\documentclass[a4paper,fleqn]{cas-dc}

\usepackage[authoryear]{natbib}
\usepackage{soul}
\sethlcolor{white}
\usepackage{ulem}
\usepackage{mdframed}

\def\tsc#1{\csdef{#1}{\textsc{\lowercase{#1}}\xspace}}
\tsc{WGM}
\tsc{QE}
\tsc{EP}
\tsc{PMS}
\tsc{BEC}
\tsc{DE}

\let\printorcid\relax

\begin{document}
\let\WriteBookmarks\relax
\def\floatpagepagefraction{1}
\def\textpagefraction{.001}
\shorttitle{3DGR-CT}
\shortauthors{Yingtai Li et~al.}

\title [mode = title]{3DGR-CT: Sparse-View CT Reconstruction with a 3D Gaussian Representation}                      
\tnotemark[1]

\tnotetext[1]{This work is supported by Natural Science Foundation of China under Grant 62271465.}

\author[1,2]{Yingtai Li}
\author[1,2]{Xueming Fu}
\author[1,2]{Han Li}
\author[1,2]{Shang Zhao}
\author[1,2]{Ruiyang Jin}
\author[1,2,3,4]{S. Kevin Zhou\cormark[1]}

\cortext[cor1]{Corresponding author: S. Kevin Zhou (e-mail: skevinzhou@ustc.edu.cn)}

\affiliation[1]{organization={School of Biomedical Engineering, Division of Life Sciences and Medicine, University of Science and Technology of China (USTC)},
                city={Hefei},
                postcode={230026}, 
                state={Anhui},
                country={China}}

\affiliation[2]{organization={China and Center for Medical Imaging, Robotics, Analytic Computing \& Learning (MIRACLE), Suzhou Institute for Advance Research, USTC},
                city={Suzhou},
                postcode={215123}, 
                state={Jiangsu},
                country={China}}

\affiliation[3]{organization={Key Laboratory of Precision and Intelligent Chemistry, USTC},
                city={Hefei},
                postcode={230026}, 
                state={Anhui}, 
                country={China}}

\affiliation[4]{organization={Key Laboratory of Intelligent Information Processing of Chinese Academy of Sciences (CAS), Institute of Computing Technology, CAS},
                city={Beijing},
                postcode={100190}, 
                country={China}}

\begin{abstract}
Sparse-view computed tomography (CT) reduces radiation exposure by acquiring fewer projections, making it a valuable tool in clinical scenarios where low-dose radiation is essential. However, this often results in increased noise and artifacts due to limited data.  
In this paper we propose a novel 3D Gaussian representation (3DGR) based method for sparse-view CT reconstruction. Inspired by recent success in novel view synthesis driven by 3D Gaussian splatting, we leverage the efficiency and expressiveness of 3D Gaussian representation as an alternative to implicit neural representation. To unleash the potential of 3DGR for CT imaging scenario, we propose two key innovations: (i) FBP-image-guided Guassian initialization and (ii) efficient integration with a differentiable CT projector. 
Extensive experiments and ablations on diverse datasets demonstrate the proposed 3DGR-CT consistently outperforms state-of-the-art counterpart methods, achieving higher reconstruction accuracy with faster convergence. Furthermore, we showcase the potential of 3DGR-CT for real-time physical simulation, which holds important clinical applications while challenging for implicit neural representations. 
\end{abstract}

\begin{keywords}
Sparse-View Computed Tomography \sep CT Reconstruction \sep 3D Gaussian Representation
\end{keywords}

\maketitle

\section{Introduction}

\label{sec:introduction}
Sparse-view computed tomography (CT) reduces radiation exposure by acquiring fewer projections, addressing the critical issue of high radiation exposure with conventional CT. It is particularly valuable in clinical scenarios where low-dose radiation exposure is preferred, such as preliminary disease screening, pediatric imaging, and follow-up, offering a safer alternative for patients and broadening diagnostic capabilities \citep{Task-Oriented,Patch-Wise-Metric-Learning,DDPNet}.
However, sparse-view CT encounters the typical inverse problem characterized by increased noise and artifacts \citep{evaluation-sparse-view-recon} due to the limited and incomplete data obtained from fewer projections.
To this, various reconstruction algorithms have been proposed, including analytical methods \citep{FDK-algorithm}, iterative reconstruction methods \citep{SART,total-variation,asd-pocs}, deep learning methods \citep{deep-sinogram-synthesis,lose-the-views,deep-learning-sparse-view-ct,dudotrans,Learnable-interpolation,Transformer-Iterative,X2Vision,FreeSeed,Alternating-Minimization,MEPNet}, and implicit neural representations (INRs) methods \citep{INR-in-medical,COIL,DCTR,nerp,NAF,SNAF,DIF-Net,Teeth-NIF}. 

Deep learning methods have shown considerable promise in sparse-view CT reconstruction. However, they heavily rely on large amounts of high-quality paired data, which is often unavailable. Some generative methods, such as diffusion models~\citep{DDPM, scorebased} can solve linear inverse problems without paired data for training~\citep{song2021solving, chung2022diffusion, chung2023solving}, while they may introduce biased prior from training data to reconstructed image, causing additional artifacts. Furthermore, generalization problems arise when used across different acquisition sites and conditions. 

INRs have been recognized as a promising approach to sparse-view CT reconstruction, owing to their distinct advantages such as “self-supervised” training and robustness against distribution-shift. 
Unlike learning-based methods that learn a direct mapping from projection data or low-quality reconstructed images to high-quality images, INRs, typically implemented using Multi-Layer Perceptrons (MLPs) augmented with positional encoders, learn the mapping from spatial coordinates to intensity values. The paramters of the MLP are optimized by minimizing errors in the projection space \citep{nerp}. 
This "self-supervised" learning approach aligns closely to the physical process of CT imaging, thus eliminating the need for paired projection-image data \citep{nerp,NAF} and mitigating challenges posed by distribution-shift arising from variations in imaging devices and parameters. 
However, INR methods face challenges such as high computational costs and spectral bias. They suffer from a high computational cost, as determining the CT value at each coordinate requires the involvement of all network parameters. 
Additionally, INRs exhibit a spectral bias, favoring low-frequency information over high-frequency details \citep{spectral-bias}, which hinders the accurate reconstruction of sharp boundaries.  While subsequent studies have attempted to alleviate this problem \citep{fourier-features,instant-NGP,NAF} by designing various kind of positional encodings, these encodings often introduce their own inductive biases, prone to result in over-smooth result or additional high-frequency artifacts \citep{CompactNGP,Accurate-Differential-Operators}.

\begin{figure*}[t]
    \centering
    \includegraphics[width=.8\textwidth]{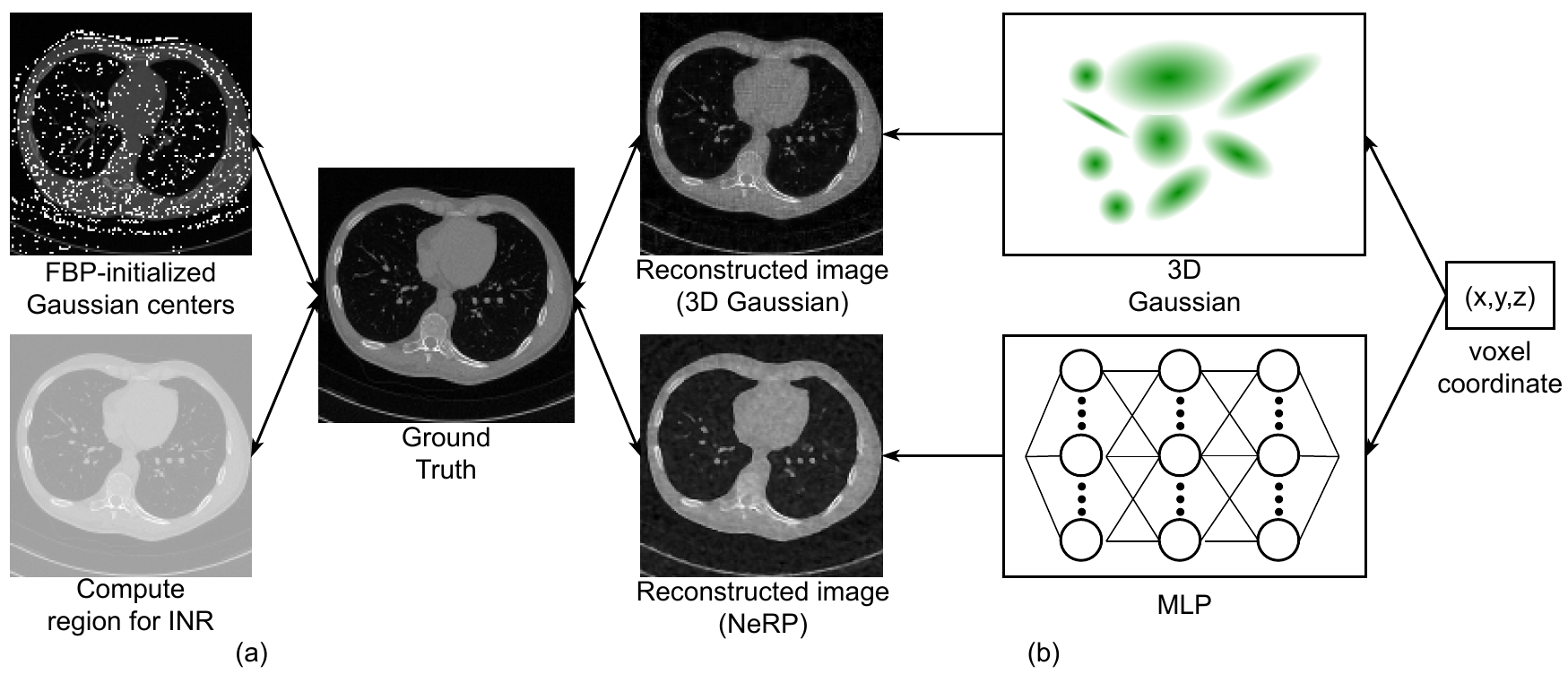}
    \caption{Comparison of 3D Gaussian Representation (3DGR) and Implicit Neural Representation (INR). (a) Prior-Informed Initialization: Our FBP-image guided initialization technique effectively avoids placing Gaussians in void regions and allocates their density (corresponds to modeling capacity) according to region complexity. 
    Computations for each voxel only involve nearby Gaussians. In contrast, INR need to compute the forward process of a fully connected network many times. (b) Adaptive Density: Neural networks treat each coordinate equally, while the density of 3D Gaussians can be adaptively adjusted during optimization through cloning and splitting operations. This dynamic allocation of modeling capacity allows for superior reconstruction of fine details. }
    \label{fig:initialization}
\end{figure*}

Recently, 3D Gaussian splatting \citep{3DGS} has emerged as a significant technical advancement in novel view synthesis, which matches or surpasses the performance of state-of-the-art NeRF methods such as Mip-NeRF360 \citep{mip-nerf360}, while drastically reducing training times and enabling real-time rendering. In contrast to INRs, 3D Gaussian representations offer a more computationally efficient alternative. This efficiency stems from their inherent local influence property: the value at any given 3D coordinate is influenced only by a small, localized subset of nearby Gaussians. 
Furthermore, 3D Gaussians can be efficiently rendered through splatting technique. 
Moreover, the number of Gaussians can be dynamically adjusted based on the local geometric complexity through the creation and destruction of Gaussians. This adaptability allows 3D Gaussians to efficiently capture high-frequency details, leading to more accurate reconstructions. We compare the two representations in Fig. ~\ref{fig:initialization}. 
The inherent flexibility of the 3D Gaussian representation unlocks possibilities for swift 3D scene editing \citep{gaussianeditor} and real-time physical simulation \citep{physgaussian}. This capability paves the way for creating dynamic ``digital twins" of reconstructed objects, opening exciting new avenues for crucial clinical applications such as surgical planning and personalized treatment strategies. 

Inspired by the success of using 3D Gaussian splatting in novel view synthesis, we propose to equip 3D Gaussian representation with modern graphics techniques for sparse-view CT reconstruction task, which marks the first such attempt in literature, to the best of our knowledge. Some pioneering works adapt 3D Guassian splatting for medical imaging senarios, including novel view synthesis \citep{GaSpCT,RadiativeGaussian,DDGS-CT,endosparse}, endoscopic tissue reconstruction \citep{endoGS}, surgical scene reconstruction \citep{LGS, EndoGSLAM} and 4D DSA rendering \citep{TOGS}. 
However, effectively initializing and optimizing Gaussian parameters within the context of CT imaging presents unique challenges, as discussed in concurrent works \citep{R2Gaussian,extremely-sparse}. 
Unlike original 3D Gaussian splatting, where Gaussian centers are initialized from point clouds obtained through Structure-from-Motion (SfM) pipelines, CT imaging provides precise knowledge of radiation source and detector positions, rendering SfM-based initialization unnecessary and potentially suboptimal. Moreover, the splatting technique employed in \citep{3DGS},  which sums projected Gaussians on a 2D plane rather than in 3D space, hinders the accurate recovery of 3D CT images. The affine approximation used to project 3D Gaussians into 2D Gaussians also introduces errors. While these inaccuracies might be acceptable in entertainment applications, precise CT values are crucial for numerous clinical applications, such as dose calculation~\citep{dose-calculation} and prediction~\citep{dose-prediction}. 

In this paper, we propose a 3D Gaussian representation-based CT reconstruction method, termed \textbf{3DGR-CT}, for sparse-view CT reconstruction. 
Specifically, 3DGR-CT has two steps tailored for CT imaging scenario: (i) FBP-image-guided Gaussian initialization. We design our method to effectively leverage prior image information by initializing Gaussians based on the FBP-reconstructed image. This strategy avoids placing Gaussians in void regions and effectively allocates modeling capacity based on regional complexity. (ii) Efficient integration with a differentiable CT projector. To overcome the inapplicability of the original splatting technique for CT imaging, we adopt a NeRF-style volumetric rendering approach, voxelize Gaussians with a customized CUDA kernel, and equip them with a differentiable CT projector. 

\begin{figure*}[t]
    \centering
    \includegraphics[width=.8\textwidth]{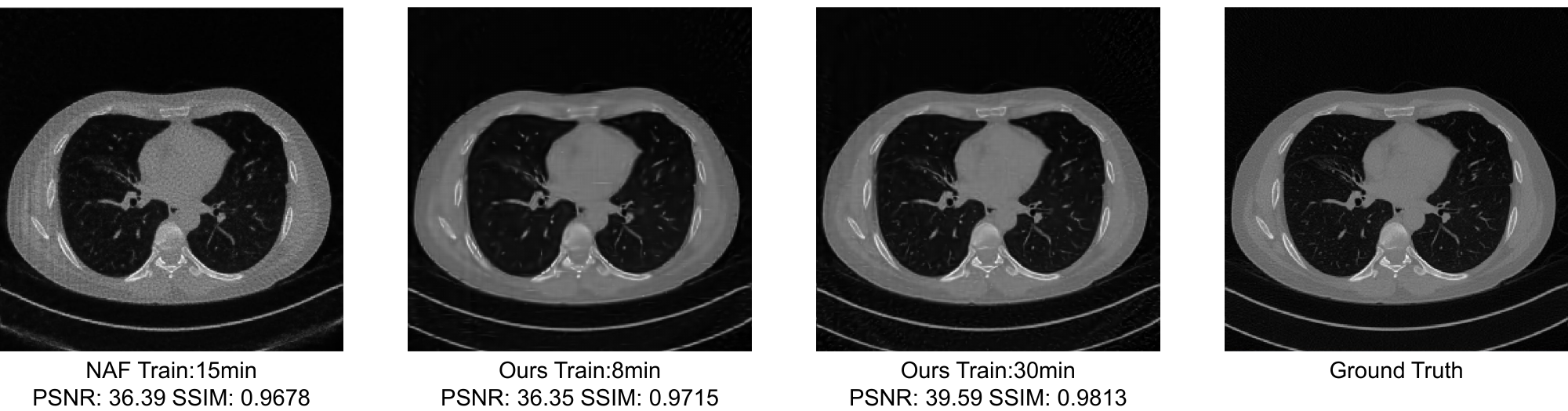}
    \caption{Our method achieves similar quantitative performance with state-of-the-art INR based method with only half of its time, and continue to arrive at a much better visual and quantitative result given more time.}
    \label{fig:speed-performance}
\end{figure*}

By comparing differences in the projection space, we can accurately update the parameters of the Gaussians and recover precise CT values. Such a process inherits the advantages of INRs, including "self-supervised" training and robustness to distribution-shift, while surpassing INRs in terms of both reconstruction quality and convergence speed. As demonstrated in Fig.~\ref{fig:speed-performance}, 3DGR-CT outperforms state-of-the-art INR methods in both aspects. We conduct extensive experiments on diverse datasets encompassing chest, abdomen, foot, jaw, head and vertebrae CT scans, where 3DGR-CT consistently outperforms existing methods. 
Furthermore, we showcase the feasibility of performing physical simulations on coronary artery reconstructed using the proposed 3DGR-CT, a task that poses significant challenges for INR-based representations.

In summary, our contributions are threefold:
\begin{itemize}
    \item \textbf{Equipping 3D Gaussian Representation with Modern Graphics Techniques:} We propose a method that leverages 3D Gaussian representations as a flexible and efficient approach for reconstructing sparse-view CT images. By equipping it with adpative density control, our approach achieves highly competitive reconstruction accuracy and significantly improved convergence speed compared to implicit neural representation (INR) methods.

    \item \textbf{Tailored Initialization and Optimization Strategies:} We develop customized initialization and optimization techniques specifically designed for CT imaging, powered by customized CUDA kernel. These innovations circumvent the limitations of Structure-from-Motion (SfM) based initialization and avoid the inapplicability of original splatting technique for CT imaging. 

    \item \textbf{Comprehensive Evaluation and Demonstration of Potential:} We showcase the superiority of 3DGR-CT through rigorous evaluations across a wide range of datasets and conduct thorough ablation studies. Moreover, we highlight the potential of this representation for advanced applications like physical simulation, which are difficult to achieve with implicit representations while important for many clinical applications.
\end{itemize}

\section{Preliminaries and Related Work}
\subsection{3D Gaussian Splatting Preliminaries}
Kerbl et.al.\citep{3DGS} present a novel 3D Gaussian splatting approach for real-time radiance field rendering, which achieves state-of-the-art visual quality while maintaining competitive training times. The key to this performance is a new 3D Gaussian scene representation coupled with a real-time differentiable renderer, which offers a significant speedup for both scene optimization and novel view synthesis.
The core of this approach is a differentiable and flexible representation using 3D Gaussians, which can be efficiently rendered through a splatting technique. 

Each 3D Gaussian is defined by several parameters:
\begin{itemize}
    \item Position ($\mu$): The mean position of the Gaussian in 3D space.
    \item Opacity ($\alpha$): The transparency level of the Gaussian.
    \item Covariance ($\Sigma$): The covariance matrix defining the shape and orientation of the Gaussian.
    \item Spherical Harmonic (SH) Coefficients: These coefficients represent the color $c$ and its directional dependencies.
\end{itemize}

The 3D Gaussian is mathematically expressed as: 
\begin{equation}
    G(x) = \exp\left(-\frac{1}{2}(\bf{x}-\bf{\mu})^T\Sigma^{-1}(\bf{x}-\bf{\mu})\right),
\end{equation}
where $\mathbf{x}$ represents a point in 3D space.

During rendering, these 3D Gaussians are projected onto a 2D image plane, forming ``splats." 
The covariance matrix of the projected 2D Gaussian, denoted as $\Sigma^{'}$, is obtained through the following transformation~\citep{zwicker2001ewa}: 
\begin{equation}
\label{equ:splatting}
    \Sigma^{'} = JW \Sigma W^TJ^T,
\end{equation}
where $J$ is the Jacobain of the affine approximation of the projective transformation, and $W$ is the view transformation. Employing the affine approximation is generally adequate in most cases. However, it can introduce noticeable errors when dealing with large Gaussians or Gaussians located near the image boundaries. 

The color $C$ of a pixel is determined by $\alpha$-blending the contributions of all $N$ splats that influence that pixel, considering their order:
\begin{equation}
    C = \Sigma_{i\in N} c_i\alpha_i\Pi_{j=1}^{i-1} (1-\alpha_j).
\end{equation}
Here, $c_i$ represents the color of the i-th Gaussian splat at the pixel location. 

The optimization process iteratively refines the Gaussians' parameters to minimize the difference between the rendered views and the ground truth projections. An adaptive density control mechanism ensures accurate scene representation by adding or removing Gaussians based on their view-space positional gradients and a opacity threshold.

\subsection{Related Work}

\textbf{Basis Functions in CT Reconstruction}

Basis functions play a fundamental role in computerized tomography reconstruction \citep{herman2009fundamentals}. The choice and properties of basis functions significantly impact both reconstruction quality and computational efficiency \citep{7145450}. Traditional approaches often use fixed basis functions like pixels/voxels or generalized kaiser-bessel window functions~(blobs). While our work primarily focuses on Gaussian functions, the proposed framework is flexible and can be extended to incorporate other basis functions, potentially combining their respective advantages.

\textbf{Radial Basis Functions}

Radial Basis Function (RBF) methods represent a signal as a weighted combination of radially symmetric basis functions centered at specific points. While RBF networks have shown promise in representing and reconstructing \citep{carr2001reconstruction}, including CT imaging, they face several limitations. Traditional RBF approaches use fixed centers and widths, which can be computationally efficient but may struggle to capture varying levels of detail. More recent adaptive RBF methods allow for parameter adjustment but often face challenges in balancing computational cost with representation accuracy.

Our proposed 3D Gaussian representation advances these foundational concepts through two key innovations: (1) an adaptive density control mechanism that dynamically optimizes both the number and parameters of Gaussians during reconstruction, effectively addressing the flexibility limitations of traditional RBFs, and (2) a CT-specific initialization strategy that leverages domain knowledge to guide the reconstruction process. This approach achieves superior reconstruction quality while maintaining computational feasibility, effectively bridging the gap between representation capability and computational efficiency.

\section{Method}
Building upon the success of 3D Gaussian splatting in rendering novel views, we explore representing CT image volumes with a set of 3D Gaussian functions (or Gaussians). 
While this may appear similar to tradional Radial Basis Function (RBF) based methods, our approach differs in several key aspects: (1) we employ an adaptive density control mechanism that dynamically adjusts the number, position, and scale of Gaussians during optimization, unlike RBF with fixed centers and widths, which enables more accurate representation of complex structures; (2) we leverage domain-specific prior knowledge through our FBP-guided initialization strategy, which is not available in traditional RBF approaches. 
In contrast to the original 3D Gaussian Splatting, we omit the opacity and spherical harmonic coefficients and redefine each Gaussian $G_i$ with $\theta_i=[\mu_i, \Sigma_i, t_i]$, where $\mu_i = (x_i, y_i, z_i)$ represent its center coordinate, $\Sigma_i$ represent the covariance matrix, and $t_i$ represent the intensity. 
The contribution of each Gaussian to a point $X = [x, y, z]$ in 3D space is given by:
\begin{equation}
    G_i(X|\theta_i) = t_i * e^{-\frac{1}{2}(X-\mu_i)^T \Sigma_i^{-1}(X-\mu_i)}.
\end{equation}

As in the original 3D Guassian splatting framework~\citep{3DGS}, we decompose the covariance matrix $\Sigma$ into a scaling matrix $S$ and a rotation matrix $R$, such that $\Sigma = RSS^TR^T$. This decomposition ensures the positive-definiteness of $\Sigma$ and allows for independent optimization of these parameters. We also represent scaling using a 3D vector $s$ and rotation using a quaternion $q$. To provide a reasonable starting point for optimization, Gaussian parameters are initialized using a Filtered Backprojection (FBP) image-guided process. 

As mentioned before, we find that the splatting technique employed in \citep{3DGS} inapplicable for reconstructing CT image volumes.  
Consequently, we adopt a NeRF-style volumetric rendering approach to synthesize projections, such an approach inherently aligns with the physical imaging process of X-ray CT. 
In practice, we discretize the Gaussians over coordinates of the image volume, and integrate them with a differentiable CT projector. This approach offers a more accurate representation of CT imaging physics and enables the precise recovery of intensity values. While volumetric rendering is slower than splatting, we leverage the localized influence property of 3D Gaussians and implement a customized CUDA kernel to accelerate computations. This optimization results in a processing speed faster than instant INR method, while simultaneously achieving superior accuracy in intensity value recovery. 

The optimization of Gaussians' parameters is guided by a loss function that minimizes the discrepancy between the rendered views and the ground truth projections. To further enhance reconstruction quality and adapt to varying levels of detail, Gaussian density is modulated using an adaptive density control mechanism. 
Once the optimization process is complete, the 3D image volume can be obtained by calculating the intensity value at each grid point. 

\begin{figure*}[t]
    \centering
    \includegraphics[width=\textwidth]{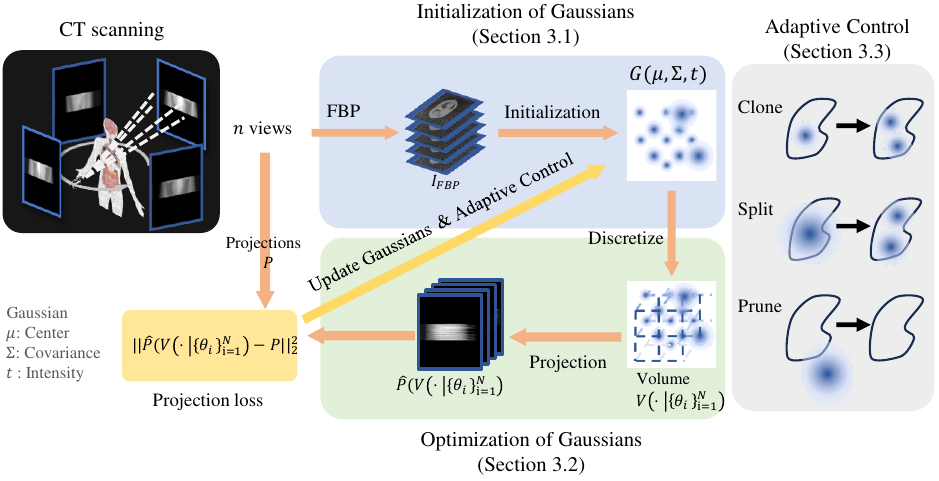}
    \caption{Overview of pipeline: Initially, projection data is acquired from various viewpoints and subsequently processed using Filtered Back Projection (FBP) to yield the FBP-reconstructed image. This reconstructed image is then utilized as a prior for initializing the 3D Gaussians. During the training phase, these 3D Gaussians are discretized into a volumetric image, which then undergoes a forward projection process. We compare the difference between these projections and the ground truth measurements, and use the gradients to update parameters of Gaussians. Ultimately, the interplay of the gradient and the Gaussians' parameters dictates the density control of the Gaussians. }
    \label{fig:pipeline}
\end{figure*}

An overview of the pipeline is in Fig. \ref{fig:pipeline}. In the following sections, we elaborate on the key steps involved: initialization of the Gaussians, updating their parameters through optimization, and conducting adaptive density control.

\subsection{FBP-Image-Guided Gaussian Initialization}
Compared to INRs, 3D Gaussians can benefit more from prior information. While \citep{3DGS} utilizes points from an SfM process for initialization, this is not ideal for CT. The known imaging geometry of the CT system makes SfM redundant and suboptimal. Instead, the FBP-reconstructed image provides a strong prior for Gaussian initialization. 

We begin by obtaining the FBP-reconstructed image $I_{FBP}(X)$ from the acquired projection data $P$. A threshold $\tau$ is applied to exclude the empty space to get ${\bar I}_{FBP}(X)$: 
\begin{equation}
    {\bar I}_{FBP}(X) = I_{FBP}(X)~\pi[I_{FBP}(X) > \tau],
\end{equation}
where $\pi[\cdot]$ represents an indicate function.

Next, we compute the gradient $\nabla {\bar I}_{FBP}$ of each voxel. The voxel coordinates ${V_i}$ are ranked by their gradient magnitude $||\nabla {\bar I}_{FBP}(V_i)||$. Voxels with very large gradients are usually affected by streaking artifacts; therefore we initialize the Gaussian centers with coordinate of medium magnitude voxels. 

For each initialized point $C_j$, we calculate the number of their neighbors $N(C_j, r)$ within a distance $r$. The covariance $\Sigma_j$ of the Gaussian at $C_j$ is initialized isotropically, represented by a scalar value $\sigma_j$ set inversely proportional to the number of neighbors. The intensity $t_j$ is set to be proportional to the intensity of the FBP-reconstructed image at $C_j$. This FBP-guided initialization differs from uniform initialization where Gaussians would be uniformly distributed in space with identical scaling and intensity values.

\begin{equation}
\begin{aligned}
        \sigma_j &= k_{\sigma} \frac{1}{N(C_j, r)}, \\ 
        t_j &= k_{t} {\bar I}_{FBP}(C_j),
\end{aligned}
\end{equation}
where $k_{\sigma}$ and $k_t$ are coefficients determining the Gaussian width and intensity, respectively. 
For isotropic Gaussians, the scaling parameter $s$ is initialized and restricted to be the same for all directions as $s=[\sigma_j, \sigma_j, \sigma_j]$, while for anisotropic Gaussians, the scaling parameters can be different in each direction during the optimization process. 
The rotation parameter $q$ is initialized to the identity rotation $q=[1,0,0,0]$. 
As we initialize the Gaussian centers using the index of the voxel grid (e.g., ranging from [0, 256] for a $256\times256\times256$ image), we normalize the initialized Gaussian centers by dividing their maximum range to map them into [0,1]. 
The intensity of CT image is also normalized to [0,1]. We utilize a sigmoid activation function to ensure that the intensity values remain within the desired range. For scaling, we employ an exponential activation function.

While INR-based methods can leverage prior information for parameter initialization, as demonstrated in NeRP \citep{nerp}, there are key differences in our approach. NeRP relies on prior images from either different phases of a 4DCT scan or previous acquisitions of the same subject - data that is not consistently available in clinical practice. In contrast, our method utilizes only the FBP-reconstructed image from the current acquisition, making it more practical for routine clinical use. Furthermore, while NeRP could theoretically use FBP-reconstructed images for initialization, our approach offers two distinct advantages: it eliminates the need for an additional fitting stage, and more importantly, our experimental results (Table \ref{tab:main-results}) demonstrate that FBP-based initialization of NeRP does not yield performance improvements. This underscores the effectiveness of our more direct, Gaussian-based approach to utilizing FBP information.

\subsection{Two-stage Gaussian Optimization}
\label{sec: optimization-of-gaussians}
To effectively optimize the parameters of our 3D Gaussian representation and accurately retrieve CT intensity values, we depart the splatting technique employed in \citep{3DGS}. Instead, we integrate Gaussians with a differentiable CT projector \citep{odl}. This integration necessitates discretizing the Gaussians onto a grid of voxels representing the CT image volume. 
However, directly computing the contribution of all Gaussians at every voxel proves computationally expensive. To address this, we restrict the computation within a 99\% ($\mu \pm 3 \sigma$) confidence interval for each Gaussian. This targeted approach significantly reduces the computational load without sacrificing accuracy. However, PyTorch’s slicing operations are limited to patches of a common size. To overcome this limitation, 
we implement a customized CUDA kernel for this computation, achieving a 10x speedup for optimization and an order of magnitude reduction in memory consumption. The contribution of all relevant Gaussians at a point $X$ in the volume is then given by: 
\begin{equation}
V(X|\{\theta_i\}_{i=1}^N) = \sum_{i : ||X-\mu_i||\leq 3\sigma_i} G_i(X|\theta_i).
\end{equation}
This localized computation significantly reduces the computational burden without compromising accuracy. 
Building upon this efficient computation strategy, we further enhance the optimization process by adopting a two-stage approach:

\subsubsection{Low-Resolution Optimization}
In the first stage, we sample a low-resolution grid, specifically $\frac{1}{2}\times\frac{1}{2}\times\frac{1}{2}$ of the original volume. This allows for rapid optimization, quickly placing the Gaussians in approximately correct positions and providing a rough initial reconstruction.

\subsubsection{Full-Resolution Refinement}
Once the initial optimization is complete, we proceed to the second stage, where we sample over all voxels of the original volume. This allows for fine-grained adjustments to the Gaussian parameters, capturing finer details and resulting in a higher-quality reconstruction.

Throughout both stages, the differentiable CT projector generates projected data $\hat{P}$ from the discretized volume. We quantify the discrepancies between these projections and the actual measurements $P$ using the $L_2$ norm. 
\begin{equation}
\centering
\min_{\{\theta_i\}_{i=1}^N} || \hat{P}[V(\cdot|\{\theta_i\}_{i=1}^N)] - P ||_2^2.
\end{equation}
By minimizing this loss function, we iteratively refine the Gaussian parameters, ultimately achieving an accurate and detailed reconstruction of the CT image volume.

\subsubsection{Customized CUDA Kernel for Gaussian Voxelization}

As PyTorch does not natively support slicing a batch of patches with different sizes, we developed a specialized CUDA kernel to compute the voxel contributions of each Gaussian within its $3\sigma$ bounding box. 
Our kernel dynamically computes each Gaussian's influence area by determining its $3\sigma$ bounding box in grid space. A work-stealing approach is used in which thread blocks atomically fetch work from a global counter. This strategy ensures balanced workload distribution even with varying sizes of Gaussian influence regions. Additionally, shared memory is utilized to cache frequently accessed Gaussian parameters (e.g., means, covariances, and scales), which dramatically reduces global memory traffic and accelerates the computation.
For thread safety, when accumulating the intensity contribution from multiple Gaussians to a single voxel, atomic additions (using \texttt{atomicAdd}) are employed. The backward pass similarly leverages these techniques to compute gradients with respect to each Gaussian parameter without introducing race conditions. The following pseudocode summarizes the forward pass of our CUDA kernel:

\begin{center}
    \resizebox{0.48\textwidth}{!}{
    \fbox{
    \begin{minipage}{0.48\textwidth}
    \vspace{1ex}
    \begin{center}\textbf{Algorithm: Gaussian Voxelization Forward Pass}\end{center}
    \vspace{1ex}
    \begin{tabular}{ll}
    1: & \textbf{For} all Gaussian $G_i$ in parallel: \\
    2: & \hspace{1em} Load parameters into shared memory \\
    3: & \hspace{1em} Compute grid cell influence region \\
        & \hspace{5em} based on Gaussian scale \\
    4: & \hspace{1em} \textbf{For} all grid points $\mathbf{x}$ in influence region: \\
    5: & \hspace{2em} Compute distance vector $\mathbf{d} = \mathbf{x} - \boldsymbol{\mu}_i$ \\
    6: & \hspace{2em} Compute Mahalanobis distance $D^2 = \mathbf{d}^T\boldsymbol{\Sigma}_i^{-1}\mathbf{d}$ \\
    7: & \hspace{2em} Compute intensity contribution \\
       & \hspace{5em}  $t_i(\mathbf{x}) = t_i \exp(-\frac{1}{2}D^2)$ \\
    8: & \hspace{2em} \texttt{AtomicAdd}($V(X|\{\theta_i\}_{i=1}^N)$, $t_i(\mathbf{x})$) \\
    \end{tabular}
    \vspace{1ex}
    \end{minipage}
    }
    }
\end{center}

\begin{center}
  \resizebox{0.48\textwidth}{!}{
  \fbox{
  \begin{minipage}{0.48\textwidth}
  \vspace{1ex}
  \begin{center}\textbf{Algorithm: Gaussian Voxelization Backward Pass}\end{center}
  \vspace{1ex}
  
  \begin{tabular}{ll}
  1: & \textbf{For} all Gaussian $G_i$ in parallel: \\
  2: & \hspace{1em} Load parameters into shared memory \\
  3: & \hspace{1em} Compute grid cell influence region \\
        & \hspace{5em} based on Gaussian scale \\
  4: & \hspace{1em} \textbf{For} all grid points $\mathbf{x}$ in influence region: \\
  5: & \hspace{2em} Compute distance vector $\mathbf{d} = \mathbf{x} - \boldsymbol{\mu}_i$ \\
  6: & \hspace{2em} Compute Mahalanobis distance $D^2 = \mathbf{d}^T\boldsymbol{\Sigma}_i^{-1}\mathbf{d}$ \\
  7: & \hspace{2em} Compute intensity value $t_i(\mathbf{x}) = t_i \exp(-\frac{1}{2}D^2)$ \\
  8: & \hspace{2em} Get gradient from output $\nabla_{\text{out}}V(X|\{\theta_i\}_{i=1}^N)$ \\
  9: & \hspace{2em} $\nabla_{t_i} = \exp(-\frac{1}{2}D^2) \cdot \nabla_{\text{out}}V_\theta$ \\
  10: & \hspace{2em} \texttt{AtomicAdd}($\nabla t_i$, $\nabla_{t_i}$) \\
  11: & \hspace{2em} $\nabla_{\boldsymbol{\mu}_i} = t_i(\mathbf{x}) \cdot \nabla_{\text{out}}V_\theta \cdot \boldsymbol{\Sigma}_i^{-1}\mathbf{d}$ \\
  12: & \hspace{2em} \texttt{AtomicAdd}($\nabla\boldsymbol{\mu}_i$, $\nabla_{\boldsymbol{\mu}_i}$) \\
  13: & \hspace{2em} $\nabla_{\boldsymbol{\Sigma}_i^{-1}} = -0.5 \cdot t_i(\mathbf{x}) \cdot \nabla_{\text{out}}V_\theta \cdot \mathbf{d}\mathbf{d}^T$ \\
  14: & \hspace{2em} \texttt{AtomicAdd}($\nabla\boldsymbol{\Sigma}_i^{-1}$, $\nabla_{\boldsymbol{\Sigma}_i^{-1}}$) \\
  \end{tabular}
  \vspace{1ex}
  \end{minipage}
  }
  }
\end{center}

\subsubsection{Differentiable CT Projector and Gradient Propagation}

The differentiable CT projector is implemented to simulate the forward X-ray projection process. Let

\begin{align}
p_\theta = \mathcal{P}(V_\theta),
\end{align}

where $p_\theta$ represents the simulated projection data, $V_\theta$ is the volume assembled from the 3D Gaussians with parameters $\theta$ (short for $V(X|\{\theta_i\}_{i=1}^N)$), and $\mathcal{P}$ denotes the forward projection operator (implemented using the ODL library).

Gradients are propagated through the projection operator by applying the chain rule:

\begin{align}
\frac{\partial \mathcal{L}}{\partial \theta} = \frac{\partial \mathcal{L}}{\partial p_\theta} \cdot \frac{\partial p_\theta}{\partial V_\theta} \cdot \frac{\partial V_\theta}{\partial \theta},
\end{align}

where $\mathcal{L}$ denotes the loss function computed between the simulated and true projection data.

The derivatives of the reconstructed intensity $V_\theta$ with respect to the Gaussian parameters are explicitly derived as:
\begin{align}
\frac{\partial V_\theta}{\partial I_i} &= \exp(\left(-\frac{1}{2}D^2\right) \\
&=\exp\left(-\frac{1}{2}(\mathbf{x} - \boldsymbol{\mu}_i)^T\boldsymbol{\Sigma}_i^{-1}(\mathbf{x} - \boldsymbol{\mu}_i)\right),\\[1ex]
\frac{\partial V_\theta}{\partial \boldsymbol{\mu}_i} &= t_i \exp\left(-\frac{1}{2}D^2\right) \boldsymbol{\Sigma}_i^{-1}(\mathbf{x} - \boldsymbol{\mu}_i),\\[1ex]
\frac{\partial V_\theta}{\partial \boldsymbol{\Sigma}_i^{-1}} &= -\frac{1}{2}t_i \exp\left(-\frac{1}{2}D^2\right)(\mathbf{x} - \boldsymbol{\mu}_i)(\mathbf{x} - \boldsymbol{\mu}_i)^T.
\end{align}

By chaining the volume renderer (CUDA kernel) and the projection operator in a fully differentiable manner, gradients from the projection loss are backpropagated to the 3D Gaussian parameters. This end-to-end differentiation allows optimization directly from the measured projection data.

\subsection{Adaptive Density Control}
\label{sec:density-control}
Following the original 3D Gaussian splatting \citep{3DGS}, we enhance overall accuracy and reconstruction quality through adaptively controlling the density of Gaussians. Different from the original version, we consider the gradients of $\mu$ in world space rather than view space, providing a more accurate feedback of density requirements within the 3D volume. This adaptive control mechanism automatically determines the optimal number of Gaussians based on image complexity, requiring no manual intervention during reconstruction. 
This control mechanism comprises three strategies: 
\begin{itemize}
    \item \textbf{Cloning}: In under-reconstructed areas, characterized by high gradients and small Gaussian scales, we clone the relevant Gaussian. This increases the representational capacity in these regions, allowing for finer details to be captured.
    \item \textbf{Splitting}: Conversely, in over-reconstructed areas, identified by high gradients and large Gaussian scales, we split the relevant Gaussians. This creates multiple smaller Gaussians, enabling a more nuanced representation of local features and preventing over-smoothing. 
    \item \textbf{Pruning}: Finally, to maintain computational efficiency, we prune Gaussians with near-zero intensity, as they have a negligible impact on the final rendered image.
\end{itemize}

When cloning a Gaussian, the new Gaussian inherits the original scale. One copy remains stationary, while the other inherits the gradient information and moves along this gradient during subsequent optimization. 
When splitting a Gaussian, we reduce the scale of the resulting Gaussians by a factor of 0.8. Additionally, the intensity of each resulting Gaussian is halved. This ensures that the overall density in the region remains consistent while allowing for a more detailed representation. The positions of the split Gaussians are initialized by treating the original 3D Gaussian as a Probability Density Function (PDF) and sampling from it. This ensures that the split Gaussians effectively cover the space originally occupied by the parent Gaussian.

The parameters in our method are physically motivated rather than ad-hoc. For computational efficiency, we truncate Gaussians at 3 standard deviations, following the well-established statistical principle that 99.7\% of a Gaussian's mass lies within this range. The splitting scale factor of 0.8, the intensity halving and the PDF-based sampling ensures statistically sound coverage of the original volume. These parameters remain fixed across different datasets, demonstrating their robustness in practical applications.

\begin{figure*}[t]
    \centering
    \includegraphics[width=\textwidth]{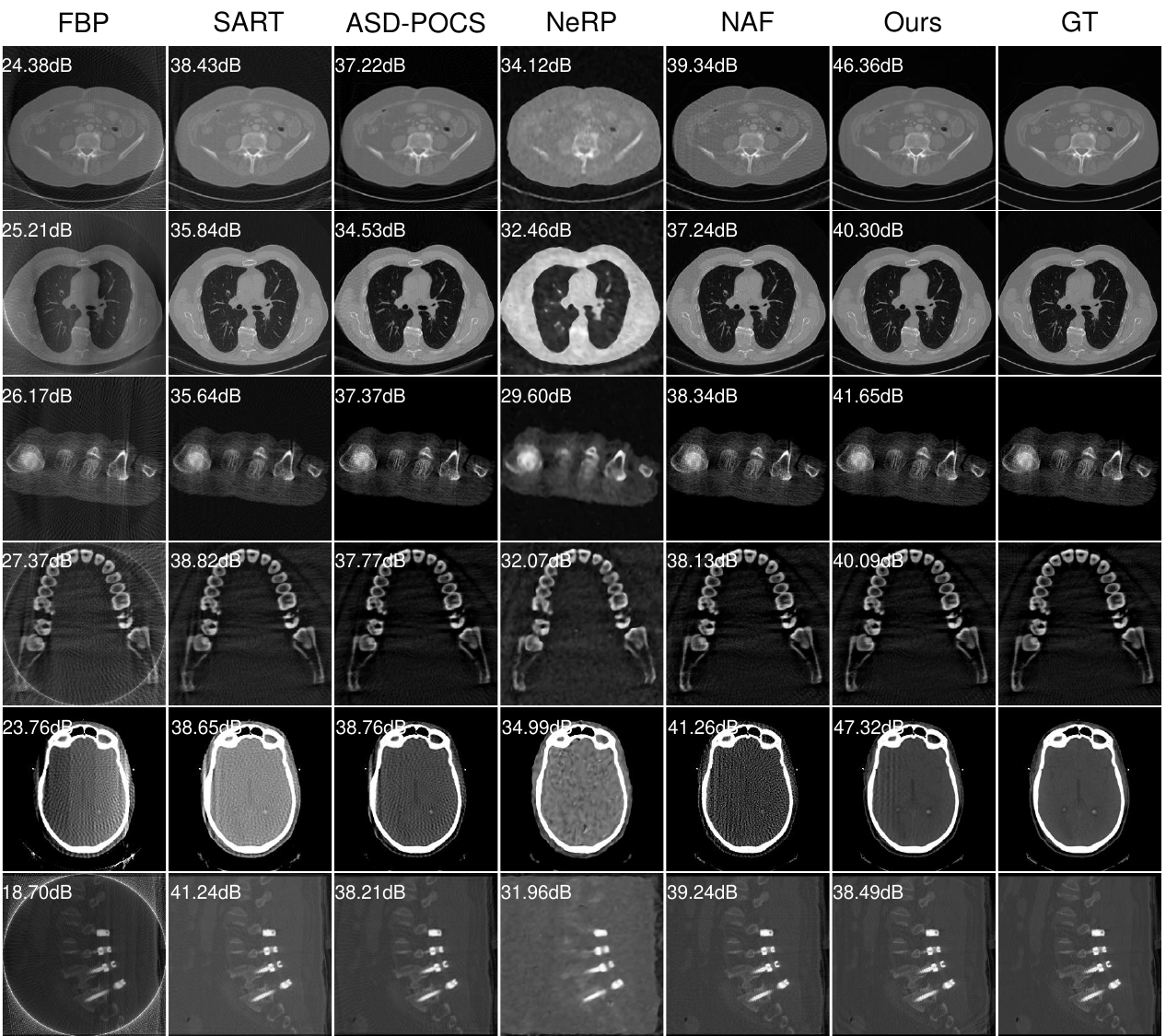}
    \caption{Visualization of reconstructed images. The window level and width are optimized for visualizing brain tissue, while other anatomical structures are displayed using standard CT value ranges. Observe that NeRP, without prior information, generates blurry reconstructions lacking intricate details. NAF, on the other hand, is prone to introducing additional noise and artifacts. Compared to INRs, our proposed 3DGR provides cleaner results in empty regions and better high-frequency details like the airways. The visual difference is especially prominent in the \textbf{abdomen}, \textbf{chest}, and \textbf{head} regions.}
    \label{fig:main-results}
\end{figure*}

\begin{figure*}[t]
    \centering
    \includegraphics[width=\textwidth]{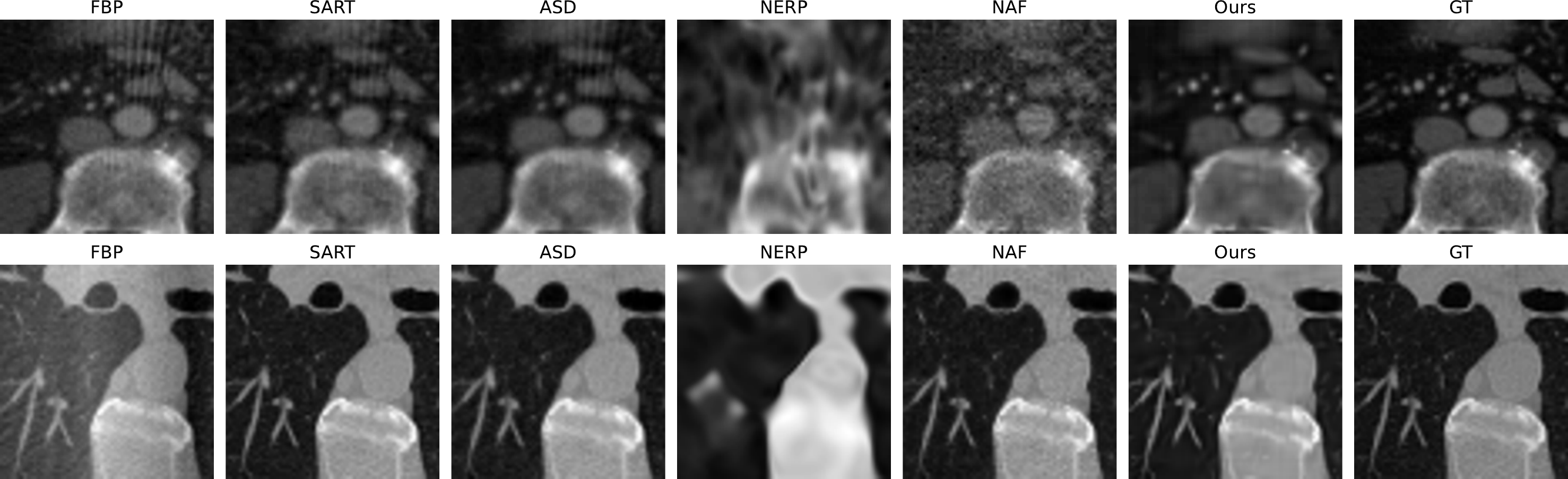}
    \caption{Zoomed-in comparison of critical anatomical structures.}
    \label{fig:zoomed-in}
\end{figure*}

\section{Experiments}
\subsection{Experimental Setup}
\textbf{Datasets} 
To validate the reconstruction performance of our method, we utilize a diverse set of CT scan datasets encompassing various anatomical regions. Specifically, we employ:
\begin{itemize}
    \item Chest and abdomen CT: AAPM-Mayo dataset \citep{AAPMDataset}
    \item Foot and jaw CT: Open Scientific Visualization datasets \citep{Klacansky2022Open}
    \item Vertebrae CT: verse2019 dataset \citep{verse-1,verse-2,verse-3}
    \item Head CT: synthRAD2023 dataset \citep{synthrad2023}
\end{itemize}

Additionally, we demonstrate the potential of physical simulation with 3D Gaussian Representation using coronary artery CT data from \citep{ASOCA1,ASOCA2}. For all datasets, we simulate a cone-beam CT acquisition process assuming 20 to 120 projections uniformly distributed over a semicircle. The scanning geometry consists of a cone-beam X-ray source and a detector with a resolution of 512 $\times$ 800 elements.

\noindent\textbf{Implementation Details} 
The image volume used in training after cropping and resizing is 
256 $\times$ 256 $\times$ 80. Both voxel coordinates and intensities are normalized to $[0, 1]$.
For the FBP reconstruction, we apply the Ram-Lak filter with frequency scaling set to 1.0. 
For training, we use Adam optimizer, with ($\beta_1$, $\beta_2$) = (0.9, 0.999), the learning rate for $\mu$ starts from 2e-3, with an exponential decay to 2e-6 by the end of the training process. We employ constant learning rates of 0.05, 0.005, and 0.001 for intensity $t$, scaling $s$, and rotation $q$ respectively. All models are trained for 15,000 iterations on an Nvidia 3090 GPU. 
Reconstructed CT images are quantitatively assessed using the structural similarity index metric (SSIM) and peak signal-to-noise ratio (PSNR).  
The 3D Gaussian results are obtained with the following setting: 
Each image volume is initialized with 50,000 Gaussians, the threshold $\tau$ used to determine background region is set to 0.05. 
The coefficients $k_{\sigma}$ and $k_I$ for initializing the scaling and intensity of each Gaussian $G_i$ is set to 0.25 and 0.15 for all body parts, which demonstrate the robustness of the hyper-parameters across different anatomical regions. We find that this initialization yields performance comparable to that of a dedicated stage for fitting the Gaussians to an FBP-reconstructed image, while avoiding extra computational overhead. Adaptive density control starts after the first 100 iterations. For cloning and splitting, the maximum permissible gradient norm for $\mu$ is set to 1e-5; For pruning, the minimum intensity of each Gaussian is set to 0.0001. 
The maximum number of Gaussians allowed during the optimization is capped at 300,000. 

\begin{table*}[t]\centering
    \caption{Quantitative Results, measured with PSNR (dB) and SSIM. Higher values indicate better performance. Average values are calculated across all reconstructed volumes. We mark the highest performance with \textcolor{red}{red}.}
    \label{tab:main-results}
    \resizebox{\textwidth}{!}{
    \begin{tabular}{l|rrrrrrrrrrrrrr}
    \hline
    \multirow{2}{*}{\textbf{Method}} 
    & \multicolumn{14}{c}{\textbf{{20-view}}} \\ \cline{2-15} 
    & \multicolumn{2}{c|}{Abdomen} & \multicolumn{2}{c|}{Chest} & \multicolumn{2}{c|}{Foot} & \multicolumn{2}{c|}{Head} & \multicolumn{2}{c|}{Jaw} & \multicolumn{2}{c|}{Vertebrae} & \multicolumn{2}{c}{Avg.} \\ 
    \hline

    FBP      &22.61 &0.651 &21.55 &0.556 &21.80 &0.399 &21.56 &0.509 &23.18 &0.521 &16.90 &0.632 &21.88 &0.588 \\
    SART     &30.77 &0.866 &29.02 &0.819 &27.29 &0.722 &28.73 &0.773 &28.76 &0.813 &\cellcolor{red!40}31.63 &\cellcolor{red!40}0.927 &29.76 &0.837 \\
    ASD-POCS &31.43 &0.908 &28.83 &0.851 &31.02 &0.921 &31.02 &0.912 &29.71 &0.856 &30.43 &0.922 &30.20 &0.883 \\
    NeRP     &30.99 &0.888 &29.41 &0.847 &28.44 &0.803 &31.73 &0.908 &27.94 &0.768 &29.03 &0.882 &30.05 &0.863 \\
    NeRP (FBP init.)       &30.67 &0.881 &29.08 &0.840 &27.35 &0.731 &28.11 &0.785 &28.11 &0.779 &28.58 &0.870 &28.65 &0.814   \\
    NAF      &31.70 &0.904 &29.77 &\cellcolor{red!40}0.866 &30.78 &0.920 &32.91 &0.942 &31.04 &0.874 &31.13 &0.904 &30.86 &0.890 \\
    Ours     &\cellcolor{red!40}34.95 &\cellcolor{red!40}0.952 &\cellcolor{red!40}29.83 &0.860 &\cellcolor{red!40}32.23 &\cellcolor{red!40}0.935 &\cellcolor{red!40}35.26 &\cellcolor{red!40}0.965 &\cellcolor{red!40}34.47 &\cellcolor{red!40}0.938 &31.16 &0.919 &\cellcolor{red!40}32.54 &\cellcolor{red!40}0.911 \\
    \hline
    
    \multirow{2}{*}{\textbf{Method}} 
    & \multicolumn{14}{c}{\textbf{{40-view}}} \\ \cline{2-15} 
    & \multicolumn{2}{c|}{Abdomen} & \multicolumn{2}{c|}{Chest} & \multicolumn{2}{c|}{Foot} & \multicolumn{2}{c|}{Head} & \multicolumn{2}{c|}{Jaw} & \multicolumn{2}{c|}{Vertebrae} & \multicolumn{2}{c}{Avg.} \\ 
    \hline

    FBP      &23.70 &0.742 &23.18 &0.694 &24.43 &0.592 &23.06 &0.646 &25.89 &0.720 &18.14 &0.711 &23.35 &0.710 \\
    SART     &33.59 &0.933 &31.77 &0.904 &30.52 &0.913 &32.97 &0.909 &32.79 &0.913 &\cellcolor{red!40}34.88 &\cellcolor{red!40}0.959 &32.70 &0.919 \\
    ASD-POCS &33.87 &0.946 &31.63 &0.911 &33.78 &0.954 &35.06 &0.960 &33.69 &0.933 &32.80 &0.945 &32.93 &0.932 \\
    NeRP     &32.67 &0.927 &31.14 &0.902 &30.04 &0.882 &34.04 &0.953 &30.92 &0.871 &31.70 &0.919 &31.87 &0.913 \\
    NeRP (FBP init.)       &32.23 &0.921 &30.21 &0.883 &29.34 &0.869 &32.77 &0.939 &31.12 &0.878 &30.84 &0.912 &31.09 &0.900  \\
    NAF      &35.64 &0.954 &32.64 &0.918 &33.85 &0.956 &38.21 &0.977 &34.67 &0.939 &33.17 &0.929 &34.28 &0.938 \\
    Ours     &\cellcolor{red!40}40.06 &\cellcolor{red!40}0.982 &\cellcolor{red!40}34.51 &\cellcolor{red!40}0.941 &\cellcolor{red!40}34.48 &\cellcolor{red!40}0.962 &\cellcolor{red!40}41.92 &\cellcolor{red!40}0.991 &\cellcolor{red!40}37.11 &\cellcolor{red!40}0.966 &34.73 &0.948 &\cellcolor{red!40}37.14 &\cellcolor{red!40}0.962 \\
    \hline
    
    \multirow{2}{*}{\textbf{Method}} 
    & \multicolumn{14}{c}{\textbf{{80-view}}} \\ \cline{2-15} 
    & \multicolumn{2}{c|}{Abdomen} & \multicolumn{2}{c|}{Chest} & \multicolumn{2}{c|}{Foot} & \multicolumn{2}{c|}{Head} & \multicolumn{2}{c|}{Jaw} & \multicolumn{2}{c|}{Vertebrae} & \multicolumn{2}{c}{Avg.} \\ 
    \hline

    FBP      &24.16 &0.789 &23.94 &0.772 &26.17 &0.737 &23.76 &0.721 &27.37 &0.824 &18.70 &0.780 &24.04 &0.778 \\
    SART     &37.28 &0.971 &35.41 &0.958 &35.64 &0.952 &38.65 &0.976 &38.82 &0.977 &\cellcolor{red!40}41.24 &\cellcolor{red!40}0.989 &36.72 &0.966 \\
    ASD-POCS &36.33 &0.971 &34.25 &0.953 &37.37 &0.980 &38.76 &0.983 &37.77 &0.971 &38.21 &0.979 &35.74 &0.965 \\
    NeRP     &33.19 &0.937 &31.36 &0.909 &29.60 &0.875 &34.99 &0.963 &32.07 &0.897 &31.96 &0.921 &32.25 &0.922 \\
    NeRP (FBP init.)       &32.43 &0.925 &30.89 &0.896 &29.57 &0.880 &31.70 &0.930 &31.87 &0.897 &32.18 &0.926 &31.44 &0.909  \\
    NAF      &39.08 &0.980 &36.18 &0.960 &38.34 &0.984 &41.26 &0.988 &38.13 &0.970 &39.24 &0.975 &37.90 &0.972 \\
    Ours     &\cellcolor{red!40}45.83 &\cellcolor{red!40}0.995 &\cellcolor{red!40}39.05 &\cellcolor{red!40}0.977 &\cellcolor{red!40}41.65 &\cellcolor{red!40}0.993 &\cellcolor{red!40}47.32 &\cellcolor{red!40}0.998 &\cellcolor{red!40}40.09 &\cellcolor{red!40}0.980 &38.49 &0.972 &\cellcolor{red!40}42.35 &\cellcolor{red!40}0.986 \\
    
    \hline
    \multirow{2}{*}{\textbf{Method}} 
    & \multicolumn{14}{c}{\textbf{{120-view}}} \\ \cline{2-15} 
    & \multicolumn{2}{c|}{Abdomen} & \multicolumn{2}{c|}{Chest} & \multicolumn{2}{c|}{Foot} & \multicolumn{2}{c|}{Head} & \multicolumn{2}{c|}{Jaw} & \multicolumn{2}{c|}{Vertebrae} & \multicolumn{2}{c}{Avg.} \\ 
    \hline

    FBP      &24.28 &0.799 &24.17 &0.794 &26.67 &0.776 &23.89 &0.736 &27.74 &0.845 &18.89 &0.807 &24.24 &0.795 \\
    SART     &39.95 &0.984 &38.47 &0.979 &39.78 &0.979 &41.57 &0.987 &\cellcolor{red!40}41.98 &\cellcolor{red!40}0.989 &\cellcolor{red!40}45.28 &\cellcolor{red!40}0.996 &39.70 &0.983 \\
    ASD-POCS &37.50 &0.979 &35.65 &0.968 &40.80 &0.991 &40.34 &0.989 &39.41 &0.980 &40.53 &0.988 &37.19 &0.976 \\
    NeRP     &33.38 &0.939 &31.49 &0.912 &30.43 &0.893 &35.76 &0.970 &31.09 &0.874 &32.64 &0.926 &32.44 &0.924 \\
    NeRP (FBP init.)       &32.43 &0.926 &30.48 &0.890 &29.47 &0.867 &30.06 &0.891 &31.56 &0.887 &32.28 &0.928 &31.05 &0.899  \\
    NAF      &41.16 &0.988 &38.71 &0.978 &41.20 &0.992 &44.33 &0.994 &39.62 &0.979 &39.02 &0.975 &40.12 &0.983 \\
    Ours     &\cellcolor{red!40}47.42 &\cellcolor{red!40}0.996 &\cellcolor{red!40}40.82 &\cellcolor{red!40}0.984 &\cellcolor{red!40}47.89 &\cellcolor{red!40}0.998 &\cellcolor{red!40}50.82 &\cellcolor{red!40}0.999 &41.58 &0.986 &40.52 &0.982 &\cellcolor{red!40}44.30 &\cellcolor{red!40}0.990 \\
    \hline
    \end{tabular}
    }
\end{table*}

\begin{figure}[h]
\centering
\includegraphics[width=0.45\textwidth]{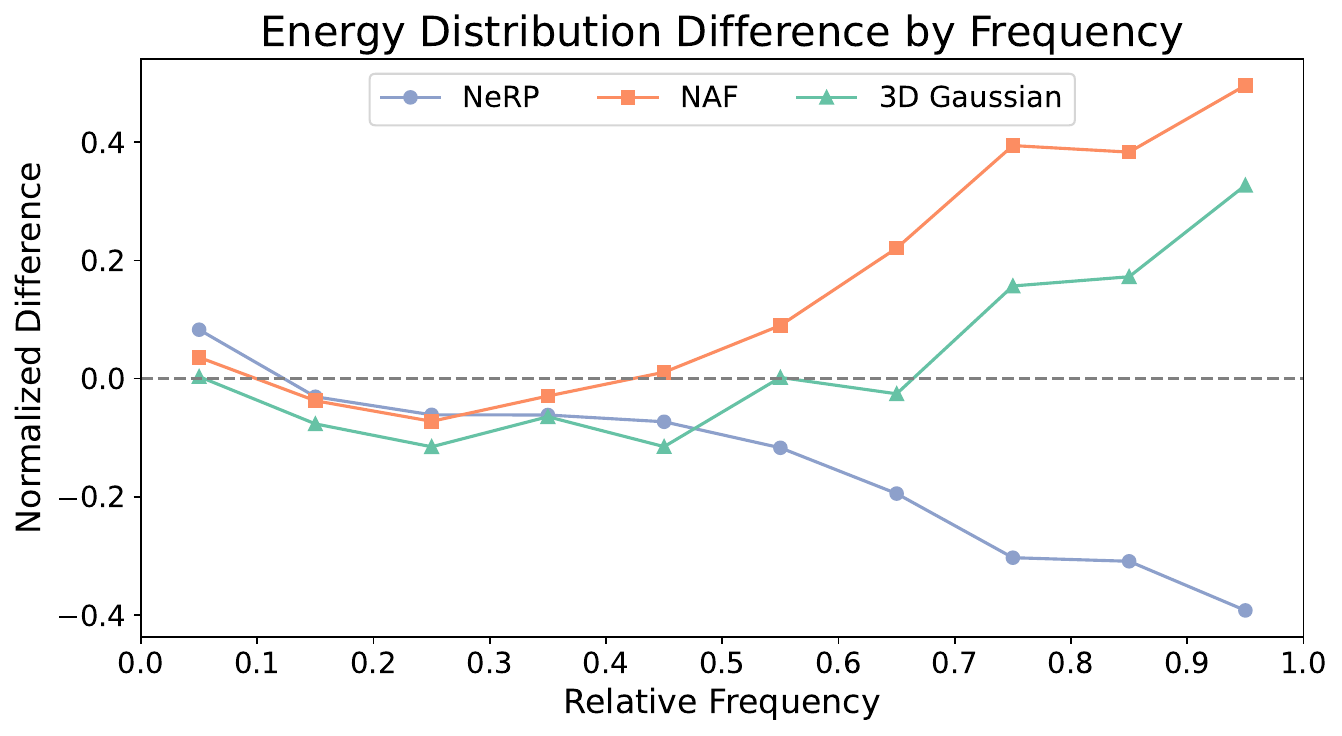}
\caption{Comparison on energy distribution difference.}
\label{fig:error-by-freq}
\end{figure}

\begin{figure}[h]
\centering
\includegraphics[width=0.45\textwidth]{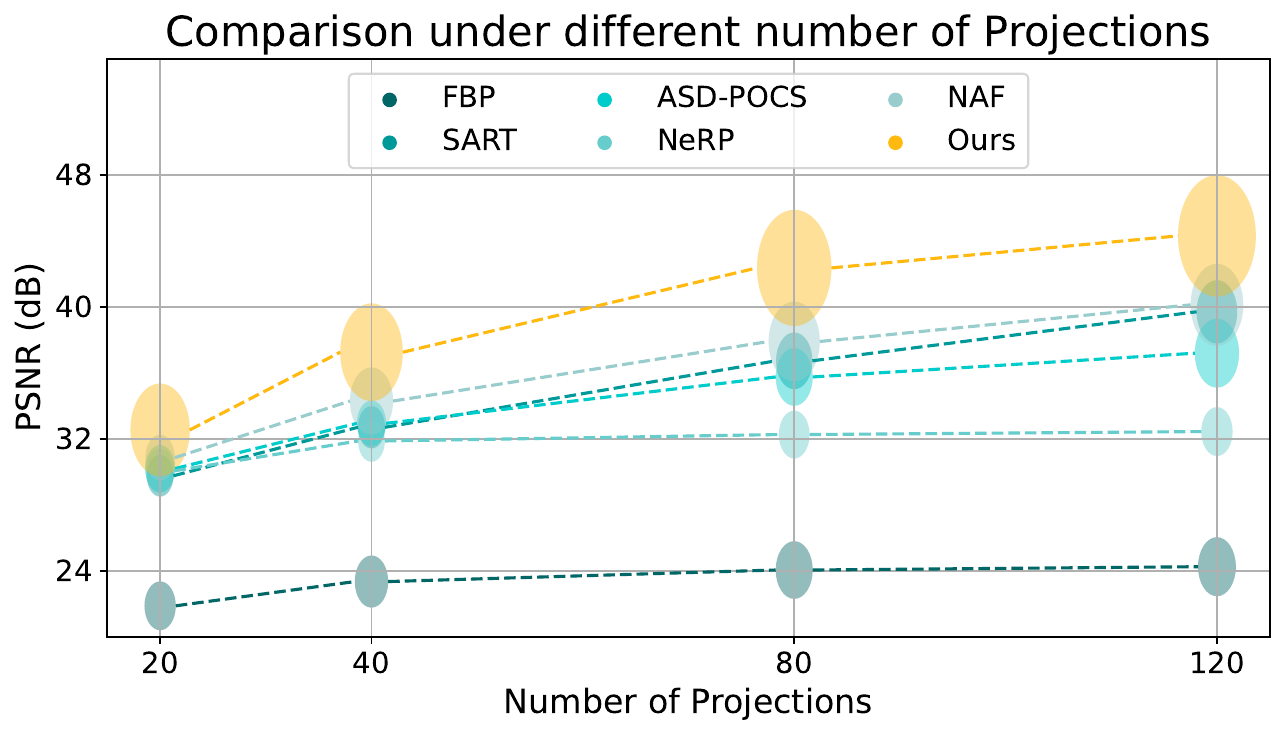}
\caption{Comparison of performance under different number of projections. The radius represents standard deviation.}
\label{fig:ablation_num_proj}
\end{figure}

\subsection{Main Results}

We compare the proposed 3D Gaussian representation with INR based methods, such as NeRP without prior \citep{nerp} and NAF \citep{NAF}. NeRP represents images using a multi-layer perceptron with a Fourier encoder, and NAF \citep{NAF} replace Fourier encoder with a multi-resolution hashing encoder, and leverages a light-weight neural network. The original NeRP paper presents two variants: one leverages a prior image (from a different phase of the respiratory cycle) to initialize the parameter of MLP, the other operates without this prior information. As such prior images are unavailable in our setting, we focus our comparison on the version of NeRP that does not rely on a prior image. For fair comparison, we have also added the results of NeRP initialized with FBP-reconstructed image. 
We also report the performance of FBP, SART \citep{SART} and ASD-POCS \citep{asd-pocs} as a baseline. 

\textbf{Quantitative} results are provided in Table~\ref{tab:main-results}. The proposed 3DGR-CT method achieves superior performance in most scenarios, usually by a large margin. The performance advantage is particularly pronounced in low-projection scenarios. The current hyperparameter settings are tuned to maximize performance in these challenging low-view conditions, and we use the same hyperparameter settings for all experiments to demonstrate the robustness. While the performance advantage may decrease slightly in certain anatomical regions as the number of views increases, the overall performance remains consistently strong and scalable. In contrast, NeRP without prior can not effectively benefit from an increased number of projection views. Initializing NeRP from FBP reconstructed image remains largely comparable with training from scratch. 
\textbf{Visual} comparisons of reconstructed images are shown in Fig.~\ref{fig:main-results}, which is obtained using 80 views. To gain a more in depth understanding, we calculate an energy distribution difference between the reconstructed image and ground truth (Fig.~\ref{fig:error-by-freq}), finding that NeRP and NAF contain too many low-frequency components. Note that over 80\% of an image's information is contained within the lowest 10\% of its frequency spectrum. NeRP struggles to capture high-frequency details, while NAF introduces additional noise and artifacts, which is also obvious from Fig.~\ref{fig:main-results}. In contrast, our proposed 3DGR-CT achieves better recovery in more frequency band. 
Performance comparison under the different numbers of projections is displayed in Fig.~\ref{fig:ablation_num_proj}. Our 3DGR-CT consistently outperform other methods. 

\begin{figure}[h]
    \centering
    \includegraphics[width=.45\textwidth]{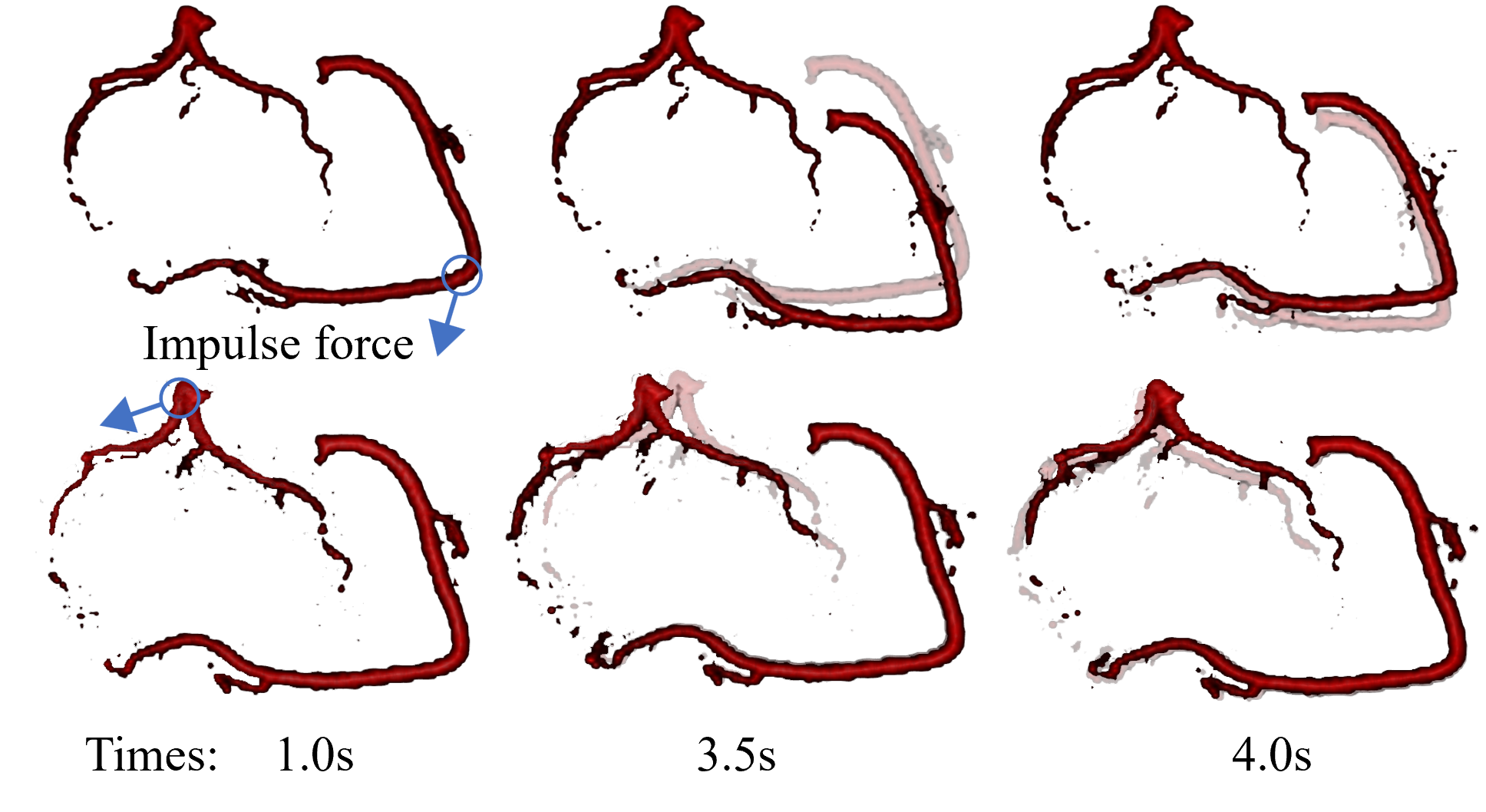}
    \caption{Demonstration of physical simulation with 3D Guassian representation. We use a blue arrow to represent the applied point and direction of the transient implusive force. The second right columns display the maximum deformation of the coronary artery. The last columns show the state at the time point when the coronary artery first begins to return to its initial position.}
    \label{fig:phys}
\end{figure}

\subsection{Demonstration of Physical Simulation}
The 3D Gaussian representation, as an explicit representation similar to meshes, facilitates various physical simulations that hold a significant importance in clinical applications, such as surgical planning and personalized treatment. In this section, we showcase a physical motion simulation perform with a reconstructed coronary artery using CT data from \citep{ASOCA1, ASOCA2}. 

We first reconstruct the coronary artery from sparse-view projections using the proposed 3DGR-CT approach, through which we obtain a set of static 3D Gaussians. 
Subsequently, we perform motion simulation using the Material Point Method (MPM), where the Gaussians are simultaneously treated as particles. These particles can record both geometric and physical information of small object regions, and they can also display the overall motion and deformation of the object through voxelization. 
For demonstration, we employ a simplified material model designed to mimic soft tissue behavior, using a “jelly” material with a Young's modulus of 2e6, a Poisson's ratio of 0.4, and a density of 200. The simulation is run in a zero-gravity setting to isolate the effects of the applied force. 
Fig.~\ref{fig:phys} shows the motion and deformation of the coronary artery after applying a transient impulsive force using the MPM simulation framework. This simulation provides a visual representation of the artery's deformation under stress, offering valuable insights for surgical planning and intervention strategies. 

To quantitatively validate the physical simulation, we performed simulations using an RTX 3090 GPU with over 50,000 particles across 100 frames. The average simulation time per frame was approximately 250 ms, with peak GPU memory consumption of 18 GB. These metrics demonstrate the feasibility of real-time simulation.

The physical simulation framework enabled by the 3D Gaussian Representation extends its utility beyond conventional diagnostic imaging. For example, in stent deployment scenarios, our simulation can provide qualitative predictions of arterial wall deformation, aiding in stent design and placement. Additionally, analyzing vessel compression dynamics could be valuable in procedures involving external forces, such as catheterization. While the current demonstration is qualitative due to simplified material models and lack of ground-truth data, it lays the groundwork for future quantitative validation and clinical application.

\subsection{Ablation Studies}
In this section we present a series of ablation studies conducted to evaluate the impact of different components of the proposed model on reconstruction quality. 

\begin{figure}[h]
\centering
\includegraphics[width=0.45\textwidth]{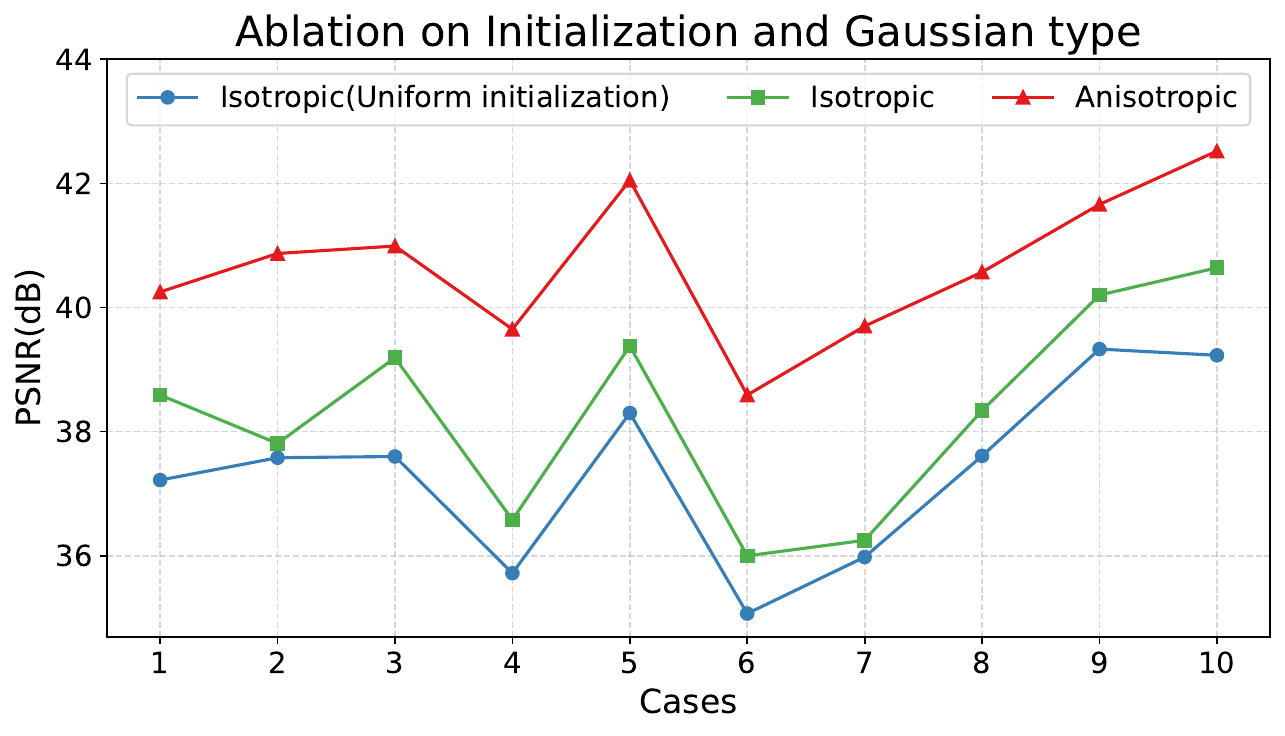}
\caption{Ablation on initialization technique and Gaussian type.}
\label{fig:ablation_ini&gaussianType}
\end{figure}

\subsubsection{Initialization of Gaussians.}
Fig. \ref{fig:ablation_ini&gaussianType} compares our proposed FBP-image-guided initialization strategy with a uniform initialization approach. For uniform initialization, we use the same threshold $\tau$ to filter the air regions and initialize Gaussians uniformly across the foreground. All Gaussisans are initialized with the same covariance and intensity. 
The results clearly demonstrate that initializing Gaussians from FBP reconstructions consistently outperforms uniform initialization by a significant margin. This highlights the importance of leveraging prior information from FBP reconstructions to achieve a more informed initial placement of Gaussians, leading to improved reconstruction quality.

We also provide a quantitative comparison with using a dedicated stage to fit Gaussians to the FBP-reconstructed image in Table \ref{tab:fbp-init}. The results show that our proposed method performs comparably to using a dedicated stage for fitting Gaussians to the FBP-reconstructed image, while avoiding the extra computational overhead.

\begin{table*}[t]\centering
    \caption{Comparison of Gaussian Initialization Strategies.}
    \label{tab:fbp-init}
    \resizebox{\textwidth}{!}{
    \begin{tabular}{l|rrrrrrrrrrrrrr}
    \hline
    \multirow{2}{*}{\textbf{Thresholds}} 
    & \multicolumn{14}{c}{\textbf{{40-view}}} \\ \cline{2-15} 
    & \multicolumn{2}{c|}{Abdomen} & \multicolumn{2}{c|}{Chest} & \multicolumn{2}{c|}{Foot} & \multicolumn{2}{c|}{Head} & \multicolumn{2}{c|}{Jaw} & \multicolumn{2}{c|}{Vertebrae} & \multicolumn{2}{c}{Avg.} \\ 
    \hline
    Fit FBP     &39.21 &0.977 &35.29 &0.952 &33.75 &0.953 &43.32 &0.992 &37.87 &0.971 &31.69 &0.913 &36.86 &0.960 \\
    Original    &40.06 &0.982 &34.51 &0.941 &34.48 &0.962 &41.92 &0.991 &37.11 &0.966 &34.73 &0.948 &37.14 &0.962 \\
  
    \hline
    \multirow{2}{*}{\textbf{Thresholds}} 
    & \multicolumn{14}{c}{\textbf{{80-view}}} \\ \cline{2-15} 
    & \multicolumn{2}{c|}{Abdomen} & \multicolumn{2}{c|}{Chest} & \multicolumn{2}{c|}{Foot} & \multicolumn{2}{c|}{Head} & \multicolumn{2}{c|}{Jaw} & \multicolumn{2}{c|}{Vertebrae} & \multicolumn{2}{c}{Avg.} \\ 
    \hline
    Fit FBP  &45.51 &0.994 &38.86 &0.975 &38.66 &0.985 &48.78 &0.998 &40.37 &0.981 &37.61 &0.965 &41.63 &0.983 \\
    Original &45.83 &0.995 &39.05 &0.977 &41.65 &0.993 &47.32 &0.998 &40.09 &0.980 &38.49 &0.972 &42.35 &0.986 \\
    \hline
    \end{tabular}
    }
  \end{table*}

\subsubsection{Isotropic vs. Anisotropic Gaussians}
Fig. \ref{fig:ablation_ini&gaussianType} also presents a comparison between isotropic and anisotropic Gaussians. Our findings indicate that anisotropic Gaussians consistently outperform their isotropic counterparts. 

\begin{figure}[h]
\centering
\includegraphics[width=0.45\textwidth]{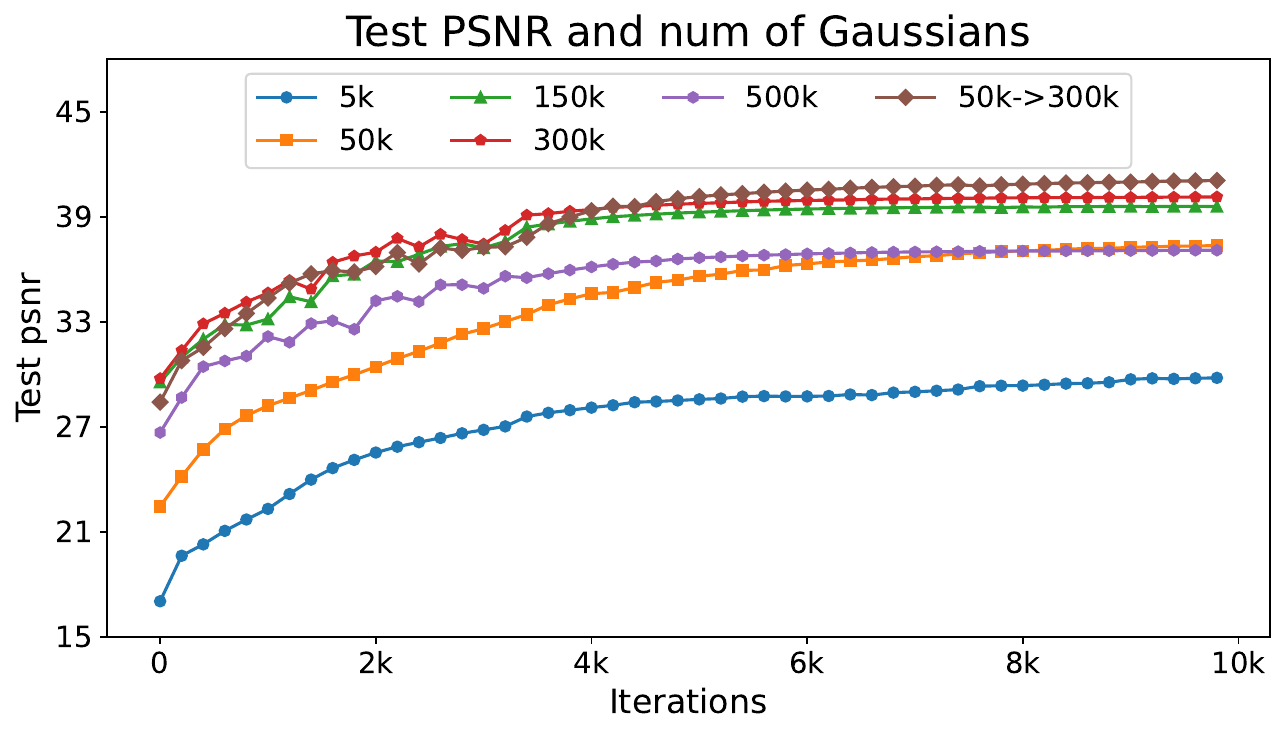}
\caption{Ablation on number of Gaussians and adaptive density control.}
\label{fig:ablation_num&denscontrol}
\end{figure}

\subsubsection{Number of Gaussian Functions.}
To understand the number of Gaussian functions and the reconstruction quality, we experiment with a range of Gaussian counts from 5,000 to 500,000. These experiments are conducted without employing adaptive density control to avoid the perturbation on Gaussian numbers. As shown in Fig. \ref{fig:ablation_num&denscontrol}, the reconstruction quality generally improves with an increasing number of Gaussians. However, using an excessively large number of Gaussians can lead to performance degradation. This is attributed to the decreased quality of initialized Gaussian centers as the number of selected points increases, making it challenging for the model to converge effectively.

\subsubsection{Effectiveness of Adaptive Density Control}
Fig. \ref{fig:ablation_num&denscontrol} also provides a comparison between reconstructions obtained with and without adaptive density control. Adaptive density control improves final reconstruction performance, outperforming direct initialization with the same number of Gaussians, which demonstrates the importance of effectively allocate the Gaussians. 

\begin{figure}[h]
\centering
\includegraphics[width=.45\textwidth]{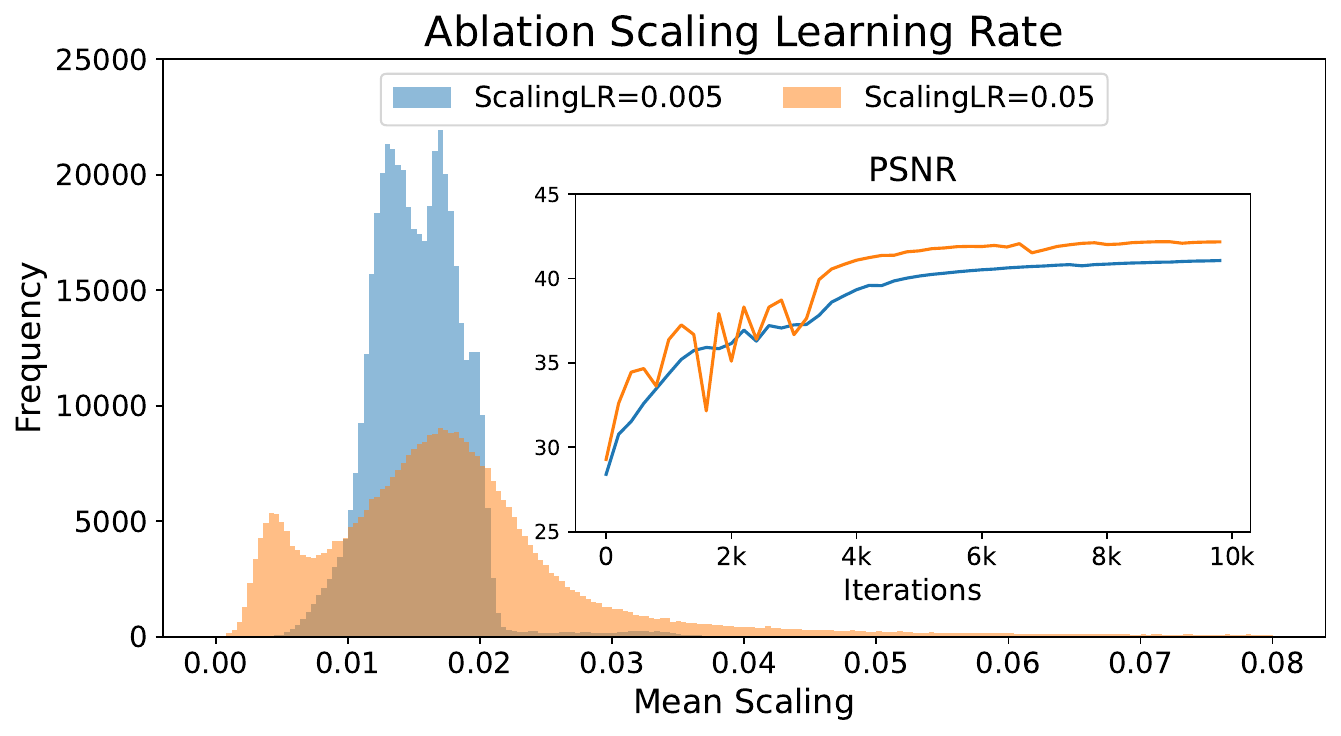}
\caption{Ablation on learning rate for scaling. }
\label{fig:ablation_lr}
\end{figure}

\subsubsection{Learning Rate for Gaussian Scale}
Figure \ref{fig:ablation_lr} illustrates the impact of the learning rate for scaling parameters on both the distribution of Gaussian scales and the PSNR of reconstructed image. 
We observe that a larger learning rate can sometimes accelerate convergence speed. However, it can also lead to numerical instability (NaN values) during optimization. Large or small learning rate for scaling generally lead to similar test performance in terms of PSNR. However, a larger learning rate tends to produce a wider distribution of Gaussian scales, with some Gaussians becoming excessively large while others become negligibly small. We find these small Gaussians contribute in a negliable fashion to the reconstruction performance, while large Gaussians significantly increases a computational demand. Empirically, we find that using a learning rate of 0.05 for scaling $s$ typically increases training time by a factor of 2-3 compared to using a learning rate of 0.005. 

\begin{figure}[h]
\centering
\includegraphics[width=.45\textwidth]{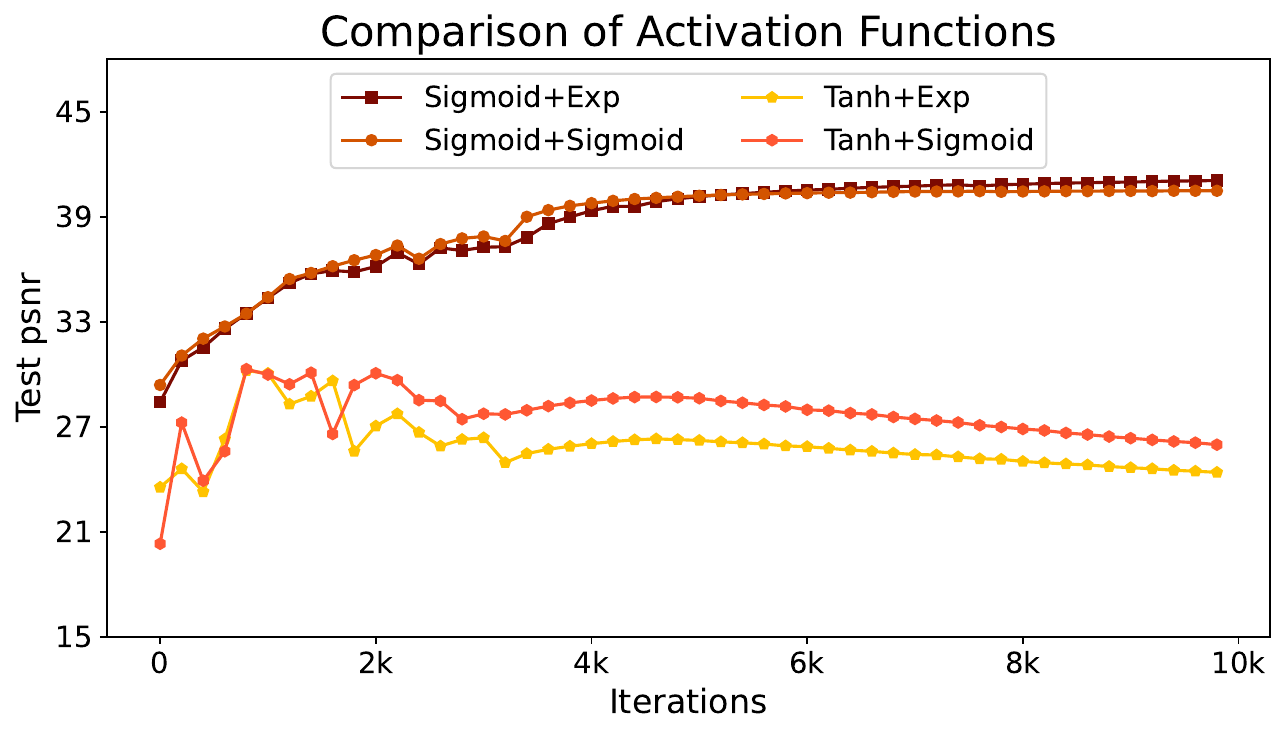}
\caption{Ablation on activation function for intensity and scaling. }
\label{fig:ablation_activation}
\end{figure}

\subsubsection{Activation Functions}
We investigate the impact of different activation functions on both intensity and scaling parameters. For intensity, we compare the performance of sigmoid and tanh activations. Our results indicate that using the tanh activation function for intensity leads to significantly worse reconstruction performance compared to using the sigmoid function. 
For scaling, we experiment with both sigmoid and exponential activation functions. We observe that both choices result in comparable performance, while the exponential activation function generally performs slightly better.

\subsubsection{Adaptive Density Control Thresholds}
We conduct an ablation study to evaluate the impact of the maximum gradient norm threshold for adaptive density control, which determines the speed of the splitting/pruning process. Table \ref{tab:ablation_density_control} shows the reconstruction performance for thresholds of 1e-4, 1e-5, and 1e-6. The results indicate that performance is relatively stable across a wide range of thresholds, demonstrating the robustness of the adaptive density control mechanism.

\begin{table*}[h]
\caption{Ablation study on adaptive density control thresholds.}
\label{tab:ablation_density_control}
\centering
\resizebox{\textwidth}{!}{
\begin{tabular}{l|rrrrrrrrrrrrrr}
\hline

\multirow{2}{*}{\textbf{Thresholds}} 
& \multicolumn{14}{c}{\textbf{{40-view}}} \\ \cline{2-15} 
& \multicolumn{2}{c|}{Abdomen} & \multicolumn{2}{c|}{Chest} & \multicolumn{2}{c|}{Foot} & \multicolumn{2}{c|}{Head} & \multicolumn{2}{c|}{Jaw} & \multicolumn{2}{c|}{Vertebrae} & \multicolumn{2}{c}{Avg.} \\ 
\hline
1e-4     &41.39 &0.987 &34.37 &0.939 &32.18 &0.940 &42.75 &0.993 &36.00 &0.952 &32.52 &0.914 &36.54 &0.954 \\
1e-5     &40.06 &0.982 &34.51 &0.941 &34.48 &0.962 &41.92 &0.991 &37.11 &0.966 &34.73 &0.948 &37.14 &0.962 \\
1e-6     &40.42 &0.983 &34.60 &0.941 &32.61 &0.945 &41.95 &0.991 &35.86 &0.950 &33.44 &0.929 &36.48 &0.957 \\
\hline
\end{tabular}
}
\end{table*}

\subsection{Comparison with Learning-based Methods}

Learning-based methods marks another important direction for CT reconstruction. We provide a basic comparison with learning-based methods by using a UNet to restore the FBP-reconstructed images, trained on 80\% of our chest and abdomen data, and a PSLD \citep{rout2023psld}(with the MAISI model \citep{guo2024maisi}) as a representative score-based generative model. The results for the remaining 20\% of the data are shown in Table \ref{tab:comparison-learning}. Our experiments reveal that pretrained diffusion models, optimized for specific resolutions, spacings, and body regions, perform poorly when applied to our test data. Strictly enforcing measurement data constraints can lead to NaN values, while relaxing these constraints can result in deviations from the ground truth. In contrast, our method adapts to different spacings, resolutions, and body regions without the need for retraining, demonstrating superior generalization. While learning-based methods could potentially benefit from more extensive training data or domain-specific fine-tuning, our findings highlight the inherent advantage of our approach in terms of generalization across varying data distributions without requiring retraining.

\begin{table}[h]
\centering
  \caption{Quantitative Results for Comparison with Learning-based Methods.}
  \label{tab:comparison-learning}
  \resizebox{.49\textwidth}{!}{
    \begin{tabular}{l|rrrrrrrr} 
      \hline
      \multirow{2}{*}{\textbf{Methods}} & \multicolumn{2}{c|}{\textbf{20-view}} & \multicolumn{2}{c|}{\textbf{40-view}} & \multicolumn{2}{c|}{\textbf{80-view}} & \multicolumn{2}{c}{\textbf{120-view}} \\
      \cline{2-9}
       & PSNR & SSIM & PSNR & SSIM & PSNR & SSIM & PSNR & SSIM \\
      \hline
      UNet       & 29.60 & 0.936 & 31.72 & 0.958 & 31.64 & 0.960 & 31.68 & 0.958 \\
      MAISI+PSLD & 20.75 & 0.527 & 21.05 & 0.533 & 21.20 & 0.554 & 21.73 & 0.545 \\
      Ours       & 34.03 & 0.931 & 37.60 & 0.962 & 42.92 & 0.987 & 44.91 & 0.989 \\
      \hline
    \end{tabular}
  }
\end{table}

\section{Validation on Real-world Data}

\begin{figure}[h]
  \centering
  \begin{minipage}{0.23\textwidth}
    \centering
    \includegraphics[width=\textwidth]{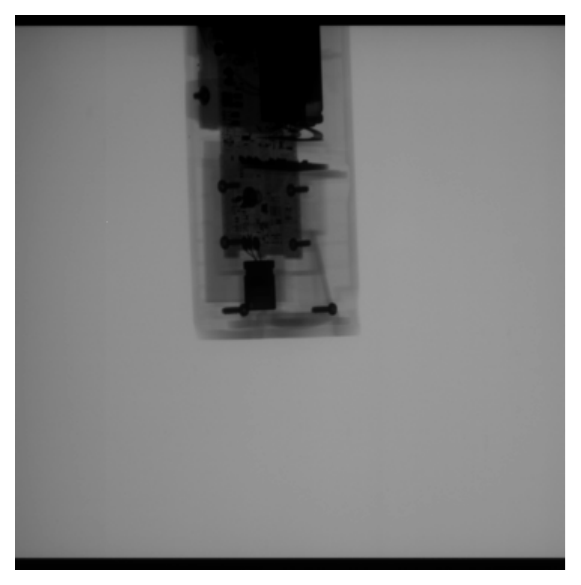}\\
    (a)
  \end{minipage}
  \hfill
  \begin{minipage}{0.23\textwidth}
    \centering
    \includegraphics[width=\textwidth]{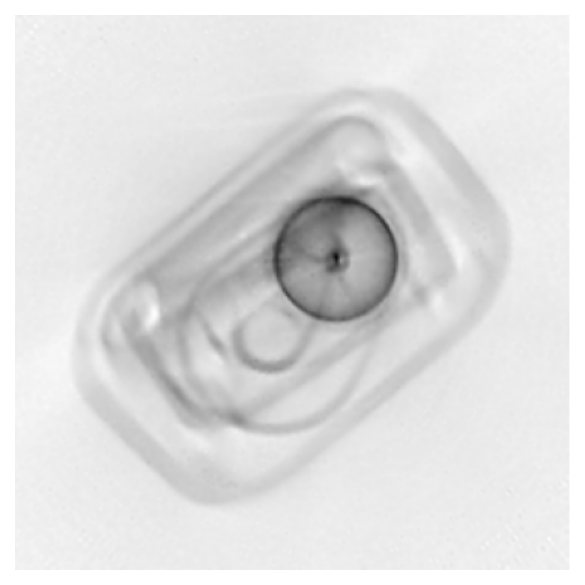}\\
    (b)
  \end{minipage}
  
  \vspace{0.1cm}
  
  \begin{minipage}{0.23\textwidth}
    \centering
    \includegraphics[width=\textwidth]{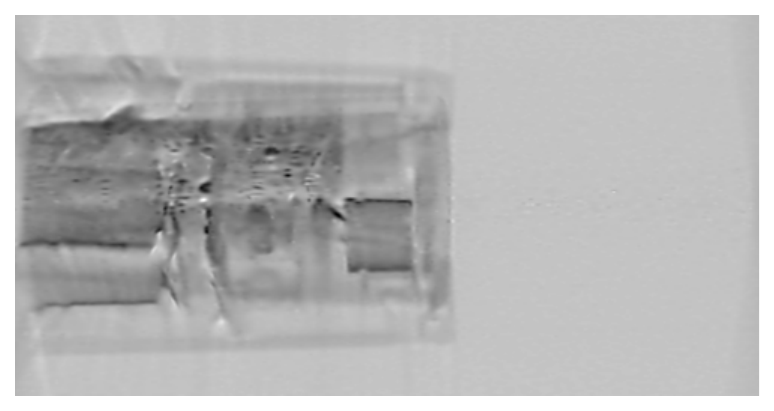}\\
    (c)
  \end{minipage}
  \hfill
  \begin{minipage}{0.23\textwidth}
    \centering
    \includegraphics[width=\textwidth]{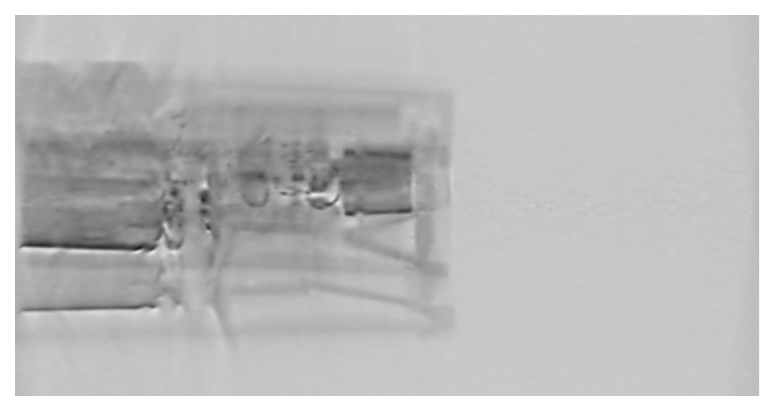}\\
    (d)
  \end{minipage}
  
  \caption{Reconstruction of real-world data. (a) Example of raw projection data. (b), (c), (d) are the reconstruction results from different views.}
  \label{fig:real-world-validation}
\end{figure}

The main experiment of our study is conducted on simulated projection data, and it is important to validate the performance on real-world data. However, CT manufacturers (e.g., Siemens, GE, Philips) employ proprietary data formats and processing pipelines. Raw projection data is typically not accessible. Scanner manufacturers often place contractual limitations on how their equipment can be operated and how resulting data can be used, further restricting the collection of experimental scanning protocols. 
As a result, We have difficulty in obtaining real-world clinical sparse-view CT data. However, we manage to obtain an industrial raw data of a CT scan, and we use it to validate the performance of our method. The results are displayed in Figure \ref{fig:real-world-validation}. As there are no ground-truth, we only provide visualization, which shows that our method works fine on real-world data.

\subsection{Performance with Different Artifacts}

While our primary focus remains on sparse-view reconstruction, we extend our evaluation to assess robustness against some common artifacts, such as scatter effect, metal artifact and motion artifact. Scatter effect is simulated by convolving projection data with a $15\times15$ blur kernel and adding it back at 10\% intensity. Metal artifacts are modeled by assigning fivefold attenuation to high-intensity regions, mimicking streaking effects from implants. Motion artifacts are introduced via random spatial shifts (up to 3 voxels) applied to 30\% of projection angles. Our method demonstrates robustness to scatter and motion artifacts, though performance degrades significantly with metal artifacts. Future work will explore dedicated correction techniques to further enhance clinical applicability.

\begin{table*}[t]\centering
    \caption{Assessment of the performance of the reconstruction method under various simulated effects.}
    \label{tab:combined}
    \resizebox{\textwidth}{!}{
    \begin{tabular}{l|rrrrrrrrrrrrrr}
    \hline
    \multirow{2}{*}{\textbf{Effect}} 
    & \multicolumn{14}{c}{\textbf{{40-view}}} \\ \cline{2-15} 
    & \multicolumn{2}{c|}{Abdomen} & \multicolumn{2}{c|}{Chest} & \multicolumn{2}{c|}{Foot} & \multicolumn{2}{c|}{Head} & \multicolumn{2}{c|}{Jaw} & \multicolumn{2}{c|}{Vertebrae} & \multicolumn{2}{c}{Avg.} \\ 
    \hline
    Original    &40.06 &0.982 &34.51 &0.941 &34.48 &0.962 &41.92 &0.991 &37.11 &0.966 &34.73 &0.948 &37.14 &0.962 \\
    \hline
    Scatter     &34.59 &0.970 &33.36 &0.947 &31.48 &0.943 &33.45 &0.984 &34.92 &0.945 &29.54 &0.916 &32.89 &0.951 \\
    \hline
    Metal       &39.42 &0.982 &32.96 &0.938 &11.07 &0.413 &15.79 &0.516 &21.54 &0.770 &14.65 &0.747 &22.57 &0.728 \\
    \hline
    Motion      &33.85 &0.948 &32.05 &0.917 &30.55 &0.909 &33.64 &0.953 &34.09 &0.937 &30.46 &0.914 &32.44 &0.930 \\
    \hline
    \multirow{2}{*}{\textbf{Effect}} 
    & \multicolumn{14}{c}{\textbf{{80-view}}} \\ \cline{2-15} 
    & \multicolumn{2}{c|}{Abdomen} & \multicolumn{2}{c|}{Chest} & \multicolumn{2}{c|}{Foot} & \multicolumn{2}{c|}{Head} & \multicolumn{2}{c|}{Jaw} & \multicolumn{2}{c|}{Vertebrae} & \multicolumn{2}{c}{Avg.} \\ 
    \hline
    Original    &45.83 &0.995 &39.05 &0.977 &41.65 &0.993 &47.32 &0.998 &40.09 &0.980 &38.49 &0.972 &42.35 &0.986 \\
    \hline
    Scatter     &35.49 &0.980 &34.16 &0.969 &34.60 &0.977 &33.81 &0.991 &37.86 &0.975 &30.52 &0.948 &34.41 &0.973 \\
    \hline
    Metal       &43.65 &0.994 &36.64 &0.976 &12.45 &0.470 &16.21 &0.624 &21.60 &0.803 &14.71 &0.843 &24.21 &0.785 \\
    \hline
    Motion      &35.08 &0.964 &33.36 &0.941 &31.15 &0.921 &34.61 &0.966 &35.37 &0.954 &32.20 &0.938 &33.63 &0.947 \\
    \hline
    \end{tabular}
    }
\end{table*}

\subsection{Comparison of interpoltation ability of Gaussian representation with INRs}
We compare the interpolation ability of Gaussian representations with INRs through following experiments: we directly fitting a low resolution image (which is obtained through downsampling the original high resolution image by a factor of $2\times2\times2$) with Gaussians, NePR and NAF seperately, then sample more dense to get an image with higher resolution. We compare the dense sampled image with the original high resolution image, the results are shown in Table \ref{tab:interpolation_ability}.

Generally NePR performs best at interpolation, which is likely due to the smooth property of neural networks. 3D Gaussian performs slightly worse than NePR. NAF, due to its use of hash encoding to discretize space into a grid where the feature at each position is obtained through a lookup table, shows worse performance at interpolation. This discretized representation results in discontinuous black regions in unsampled areas, similar to observations in \citep{wang2024hyb}. Through these experimental results, we conclude that INRs may have slightly better interpolation ability compared to 3D Gaussians if equipped with proper position encoders, but the 3D Gaussian pipeline can also use other basis functions to improve interpolation ability. Though the NeRP has the best interpolation ability, its reconstruction ability from the scratch is not the best as reconstruction goes beyond interpolation.

\begin{table*}[t]
    \caption{Interpolation ability comparison of Gaussian representations with INRs. The results are measured with PSNR(dB) and SSIM.}
    \label{tab:interpolation_ability}
    \centering
    \resizebox{\textwidth}{!}{
    \begin{tabular}{l|rrrrrrrrrrrr}
        \hline
        \multirow{2}{*}{\textbf{Method}} 
        & \multicolumn{12}{c}{\textbf{{20-view}}} \\ \cline{2-13} 
        & \multicolumn{2}{c|}{Abdomen} & \multicolumn{2}{c|}{Chest} & \multicolumn{2}{c|}{Foot} & \multicolumn{2}{c|}{Head} & \multicolumn{2}{c|}{Jaw} & \multicolumn{2}{c|}{Vertebrae}  \\ 
        \hline
    
        NeRP        &34.50 &0.950 &32.59 &0.929 &30.13 &0.889 &33.60 &0.952 &33.39 &0.919 &33.31 &0.947 \\
        NAF         &26.03 &0.759 &25.37 &0.749 &23.32 &0.824 &23.29 &0.784 &26.70 &0.821 &27.67 &0.814 \\
        3D Gaussian &34.62 &0.968 &29.41 &0.902 &27.62 &0.874 &31.18 &0.965 &34.04 &0.946 &29.06 &0.909 \\
        \hline
    \end{tabular}
    }
\end{table*}

\subsection{Comparion of computational efficiency}
We have conducted comprehensive timing experiments comparing different reconstruction methods, with results summarized in Table \ref{tab:reconstruction_time}.

\begin{table}[h]
    \caption{Reconstruction time comparison (in minutes) across different methods on various datasets.}
    \label{tab:reconstruction_time}
    \centering

    \begin{tabular}{l|ccccc}
    \hline
    Method &SART & ASD-POCS & NeRP & NAF & Ours \\
    \hline
    Time (min) & $\approx$13 & $\approx$17 & $\approx$180 & $\approx$25 & $\approx$30 \\
    \hline
    \end{tabular}

\end{table}

Our method demonstrates strong computational efficiency in practice. While modestly slower than traditional methods like SART and ASD-POCS, it achieves a significant speedup compared to NeRP (approximately 6 $\times$ faster) and maintains comparable speed with NAF. Given the potential of 3D Gaussian Representation, there are still many optimization opportunities that could further reduce computation time. All methods were implemented with CUDA acceleration and tested on the same hardware configuration for fair comparison. 

\section{Limitations and Future Work}
Despite the advantages and good performance of the proposed 3DGR-CT, it takes longer time compared with deep-learning based methods and iterative optimization methods, which could restrict its usage. 
While our current implementation is slower than traditional analytical methods, we believe this is a temporary limitation. Since the proposal of 3D Gaussian Splatting a year ago, over 2000 works have explored this representation across various applications, with many focusing on acceleration techniques \citep{accelerateGS1, accelerateGS2, accelerateGS3}. 
To address concerns regarding computational efficiency, we have explored potential hardware acceleration and algorithmic optimizations. Using an RTX 3090, our method requires approximately 30 minutes for reconstruction, which achieves a 10x speedup over naive implementations. With more advanced hardware such as an RTX 4090 or A100, we anticipate reductions in reconstruction time to approximately 18 minutes and 12 minutes, respectively. Additionally, mixed-precision computation could further enhance speed with minimal impact on quality. Algorithmic optimizations, such as combining 3D Gaussian representation with splatting techniques for faster rendering \citep{celarek2025does}, are also promising, though they may involve a trade-off between speed and quality. Techniques from recent works on compression \citep{niedermayr2024compressed} and optimized rendering pipelines \citep{girish2024eagles,wang2024adr} could be adapted to further improve efficiency. For clinical scenarios where a previous high-quality image is available \citep{nerp}, initializing the Gaussians from this image could significantly reduce reconstruction time, potentially enabling near real-time performance.
Futhermore, to achieve best performance, the hyperparameter for gaussian initialization requires manual selection, with different optimal parameters for different anatomies and resolutions. In future work, we plan to gain deeper insights of FBP-image, combining some learned priors to provide adaptive initialization and faster convergence. 

Metal artifacts are a common challenge in CT reconstruction, and our method is not immune to this issue. 
Variations in scanner hardware, such as detector resolution and X-ray spectra, can impact reconstruction performance. Differences in detector resolution may require adjusting the voxelization grid or the number of Gaussians to match the scanner's capabilities. Variations in X-ray spectra affect the FBP initialization quality but are mitigated by our ``self-supervised" optimization approach, which adapts to the specific forward model of the scanner. However, significant spectral variations may necessitate recalibration of density thresholds and scanner-specific parameter tuning. Future work could explore integrating spectral modeling or energy-dependent corrections to enhance robustness.

We believe 3D Gaussian Representation has the potential to be a successor to implicit neural representation, offering both better reconstruction quality and faster reconstruction time. As the field continues to evolve rapidly, we expect significant improvements in computational efficiency through both algorithmic advances and hardware improvements.

\section{Conclusion}
This paper presents 3DGR-CT, a novel method for sparse-view CT reconstruction based on 3D Gaussian Representation (3DGR). 
By incorporating an FBP-image guided initialization strategy and seamlessly integrating Gaussians with a differentiable CT projector, 3DGR-CT effectively unleashes the advantages of 3D Gaussian representation, overcoming limitations observed in existing INR-based methods. Our comprehensive evaluation across seven diverse datasets consistently demonstrates the superior speed and performance of 3DGR-CT compared to state-of-the-art INR-based techniques, both visually and quantitatively. 
The results highlight the potential of 3D Gaussian representation in enhancing medical imaging, offering a promising direction to reduce radiation exposure while preserving diagnostic integrity.

\bibliographystyle{cas-model2-names}

\bibliography{cas-refs}

\end{document}